\documentclass[aps,prb,twocolumn]{revtex4}  
\usepackage{amsmath,amsfonts,amssymb,latexsym}
\usepackage{hhline}
\usepackage{graphicx}
\usepackage[arrow,matrix]{xy}
\usepackage{tikz}
\usepackage{tikz-cd}
\usetikzlibrary{positioning,arrows}
\usetikzlibrary{decorations.pathmorphing}
\usetikzlibrary{decorations.markings}
\newcommand{\tp}{\otimes}
\newcommand{\ra}{\rightarrow}
\newcommand{\unit}{\mathbf{1}}
\newcommand{\ZZ}{\mathbb{Z}}

\newcommand{\mcr}{\mathcal{R}}
\newcommand{\mcz}{\mathcal{Z}}

\newcommand\be            {\begin{equation}}
\newcommand\ee            {\end{equation}}
\newcommand\ba            {\begin{aligned}}
\newcommand\ea            {\end{aligned}}
\newcommand{\mcf}{\mathcal{F}}

\newcommand{\mcc}{\mathcal{C}}
\newcommand{\mco}{\mathcal{O}}

\newcommand{\ttr}{\text{tTr}}
\usepackage{verbatim}
\newtheorem{innercustomthm}{Theorem}
\newenvironment{customthm}[1]
  {\renewcommand\theinnercustomthm{#1}\innercustomthm}
  {\endinnercustomthm}
  
\tikzset{every picture/.style=thick}
 \tikzset{
r/.style={thick,draw=black, postaction={decorate},
    decoration={markings,mark=at position .55 with {\pgftransformscale{.6}\arrow[blue]{*}}}},
l/.style={thick,draw=black, postaction={decorate},
    decoration={markings,mark=at position .5 with {\pgftransformscale{.6}\arrow[red]{*}}}},
f/.style={thick,draw=black, postaction={decorate},
    decoration={markings,mark=at position .55 with {\pgftransformscale{1.2}\arrow[black]{stealth}}}},
n/.style={decorate, draw=black,postaction={decorate},
    decoration={snake,amplitude=.3mm,segment length=1.8mm,post length=.0025mm}},
u/.style={draw=black},
arr/.style={thick,draw=black, postaction={decorate},
    decoration={markings,mark=at position 1 with {\pgftransformscale{1.1}\arrow[black]{stealth}}}},
 }
  
\begin{document}

\title{Signatures of broken parity and time-reversal symmetry in generalized string-net models}  
\author{Ethan Lake$^1$}
\email{lake@physics.utah.edu}
\author{Yong-Shi Wu$^{2,3,1}$}
\email{yswu@fudan.edu.cn}
\affiliation{$^1$Department of Physics and Astronomy, University of Utah, Salt Lake City, UT 84112, USA\\ $^2$ Key State Laboratory of Surface Physics, Department of Physics and Center for Field Theory and Particle Physics, Fudan University, Shanghai 200433, China \\ $^3$ 
Collaborative Innovation Center of Advanced Microstructures, Fudan University, Shanghai 200433, China}
\date{\today}

\begin{abstract} 

We study indicators of broken time-reversal and parity symmetries in gapped topological phases of matter. We focus on phases realized by Levin-Wen string-net models, and generalize the string-net model to describe phases which break parity and time-reversal symmetries. We do this by introducing an extra degree of freedom into the string-net graphical calculus, which takes the form of a branch cut located at each vertex of the underlying string-net lattice. We also work with string-net graphs defined on arbitrary (non-trivalent) graphs, which reveals otherwise hidden information about certain configurations of anyons in the string-net graph. Most significantly, we show that objects known as higher Frobenius-Schur indicators can provide several efficient ways to detect whether or not a given topological phase breaks parity or time-reversal symmetry.% or compute any information pertaining to the braiding data.% We consider several examples of physically relevant models in which this type of fractionalization can occur, focusing in particular on theories that host symmetry defects. 

\end{abstract}

\maketitle 

\section{Introduction}

One of the central programs in condensed matter physics in recent years has been the classification of two dimensional topologically ordered phases, which are highly entangled states of quantum matter that host quasiparticle excitations with anyonic statistics\cite{Wen04,Fradkin13}. The quintessential examples of topological phases are the fractional quantum Hall states \cite{Stern08,Wen95}, which are chiral, meaning that they break time-reversal and parity symmetry. However, a large class of topological phases that has drawn interest in recent years are so-called ``doubled'' topological phases, which are characterized by parity and time-reversal invariance \cite{Freedman04}. A promising effective model for doubled topological topological phases is the string-net model proposed by Levin and Wen\cite{Levin05}, which is an exactly soluble lattice model similar to Kitaev's toric code \cite{Kitaev06}. In string-net models, the relevant degrees of freedom are fluctuating string-like objects living on the links of a fixed two-dimensional graph, which have their origins in the mathematical framework of tensor category theory. 

String-net models can realize a large class of topological phases, namely those whose low energy effective theory is a gauge theory based on the the representations of a finite or quantum group\cite{Levin05,Hu15}. All the same, ordinary string-net models cannot realize all topological phases\cite{Lin14,Lan14}. Most significantly, ordinary string-net models cannot realize chiral phases which break either time-reversal symmetry or parity symmetry\cite{Levin05,Freedman04}. Therefore, it is important to determine the capabilities and limitations of the string-net approach, and to see whether or not the string-net idea can be generalized to encompass chiral phases. Recently, Lin and Levin have made progress in this direction by classifying the types of Abelian topological phases that certain generalized versions of string-net models can realize, and explicitly constructed variants of Abelian string-net models which break parity and time-reversal symmetries \cite{Lin14}. In order to accomplish this, they generalized the standard string-net model by expanding the graphical calculus to include extra degrees of freedom at each vertex. A natural next step in this direction is to extend their approach to encompass more general phases, including those that host non-Abelian anyons, which are of greater physical interest. % as well as extending the string-net picture to more general types of string-net lattices. 

Our principal motivation in this paper is to explore parity and time-reversal symmetry breaking in a new class of generalized string-net models. We do this by constructing invariants related to the collective response of groups of identical anyons to a type of rotation of the string-net graph, known as {\it higher Frobenius-Schur indicators}. We will show that these indicators provide two ways to test whether or not a given topological phase breaks time-reversal or parity: one way involves using only the $F$-symbols (and hence does not require solving the hexagon equations to determine the braiding data), and the other involves only the topological twists (and hence does not require solving the pentagon equations to determine the $F$-symbols). 

 When a topological phase possesses a global symmetry, its anyons can carry fractional quantum numbers of the symmetry group, which leads to the ``fractionalization'' of the symmetry action over the individual anyons in a non-trivial way \cite{Chen11,Essin13,Mesaros13b,Pollmann12}, the classic example being the fractional charge of Laughlin quasiparticles found in fractional quantum Hall liquids \cite{Laughlin83}. We demonstrate that the Frobenius-Schur indicators studied in our generalized string-net models can be derived from the response of our generalized string-net models to a certain form of orientation-preserving rotational symmetry, with the rotational symmetry fractionalizing over the individual anyons in the system.
While a great deal of theoretical work has focused on classifying the ways in which symmetry fractionalization can occur\cite{Barkeshli14,Teo15,Wen02}, most studies in this direction have focused on systems with internal symmetries as opposed to space group symmetries\cite{Barkeshli14,Hermele14,Qi15a}, although there has been some recent progress in this direction\cite{Hermele15,Qi15b,Qi15c,Song15}. Since crystal symmetries are ubiquitous in real materials, 
understanding the different ways in which space group symmetries (like the rotational symmetries realized in our model) can fractionalize is thus a key step towards identifying string-net phases in real materials.

The rest of this introduction will consist of a summary of our results, which will also serve as an outline for the paper. We begin in Section \ref{sec:review} with a brief review of the aspects of string-net diagrammatics that we will need to make use of throughout the rest of the paper.
%We implement the inclusion of symmetry defects into the theory following the general theory of $G$-crossed tensor categories as in Refs.\cite{Barkeshli14,Etingof15}, although we do not go into great detail about the theory of $G$-crossed tensor categories so as to keep our presentation accessible to a wider audience. 
In Section \ref{sec:TPS}, we review the tensor product state construction for string-net models \cite{Verstraete08,Gu09} and generalize the construction to string-net graphs of arbitrary valence. Even though all string-net graphs can be reduced to trivalent graphs, we will show that including more general structures into string-net graphs can reveal interesting information about some topological phases that is hidden when working only with trivalent graphs. Building on the approach of ref. \cite{Lin14}, we demonstrate how the tensor product state wave function possesses a natural orientation-preserving $\ZZ_n$ rotational symmetry action, which we incorporate into the theory by adding an extra branch cut into the string-net graphical calculus. We then derive a gauge-invariant quantity $\mco_n$ related to this rotational symmetry action known as the $n^{th}$ higher Frobenius-Schur indicator, which corresponds to the fractionalization of the $\ZZ_n$ symmetry action over the individual anyons in the system. We show that this type of fractionalization depends very sensitively on the structure of the underlying lattice, which could be useful in distinguishing different topological phases in an experimental setting. 

In Section \ref{sec:onsite} we demonstrate how the higher Frobenius-Schur indicators can be computed by using several different methods. One method gives $\mco_n$ in terms of the fusion data $\{[F^{abc}_d]_{ef},N^{c}_{ab},d_a\}$, and one gives $\mco_n$ in terms of the topological twists and fusion rules $\{\theta_a,N^c_{ab}\}$. In many cases, computing these indicators can allow us to quite easily determine whether or not a given topological phase breaks time-reversal or parity symmetry. 
%Since we can express the invariant $\mco_n$ entirely in terms of the fusion data, we can often determine whether or not a given set of fusion data breaks parity or time-reversal without actually needing to solve the hexagon equations or compute any information pertaining to the braiding data. 

We devote Section \ref{sec:examples} to the calculation of several examples. We begin by focusing on various bosonic SPT states and then work out some examples for a few physically relevant non-Abelian phases, namely generalized Ising theories which can be realized in bilayer quantum Hall systems\cite{Teo15,Barkeshli14}, %$\ZZ_3$ parafermions which exist in systems consisting of superconductor-Abelian quantum Hall fluid heterostructures\cite{Mong14,Fendley14,Clarke13,Hutter15}, 
and the 3-fermion state with $\ZZ_3$ symmetry which can appear on the surface of various 3D SPT states \cite{Burnell14}.

We conclude and discuss further directions in \ref{sec:discussion}. Several technical details are presented in a collection of appendices.

\section{Review of string-net diagrammatics}\label{sec:review}

We begin this section with a quick review of the diagrammatic description of the string-net model developed by Levin and Wen
\cite{Wen03,Levin05,Levin06}. The basic idea of string-net models is to use diagrams to represent operations that fuse and braid anyons. Our language is slightly different from the original language of Levin and Wen -- what they call string types we will refer to as anyonic charges (or species), and we will often refer to the oriented strings as anyon worldlines. 
%In this paper we will work with string-net graphs defined on arbitrary lattices, although all the local diagrammatic rules we will need can be written in terms of trivalent diagrams. 

The string-net labels (or the anyonic charges) are derived from a collection of objects $\mcc$, which is known in the mathematical community as a fusion category. The string-net labels $a,b,c,\dots \in \mcc$ label the different anyonic charges in the system, and mathematically they correspond to the irreducible representations of some finite or quantum group. Anyon fusion is accomplished according to the fusion rules:
\be \label{eq:fusion_rules} a \times b = \sum_{c\in \mcc} N^c_{ab} c, \ee
where the fusion rule coefficients $N^c_{ab}$ are non-negative integers, corresponding to the number of times the particle $c$ appears in the fusion product of $a\times b$. In most applications the objects $a$ and $b$ will correspond to irreducible representations of a finite or quantum group, with anyon fusion corresponding to the tensor product of representations: $a\times b = a\tp b$. If there is always one non-zero summand on the right hand side of (\ref{eq:fusion_rules}) for arbitrary $a$ and $b$, the theory represents Abelian anyons, whereas the theory describes non-Abelian anyons otherwise. We will focus only on theories that have $N^c_{ab} \in \{0,1\}$ for all $a,b,c \in \mcc$, since they encompass most models of physical interest and since the generalization to arbitrary $N^c_{ab}$ is straightforward.

In string-net diagrams, each string type $a \in \mcc$ is represented by a directed line segment, which can be interpreted as the worldline of the anyon $a$. Topological phases always contain a unique vacuum string type $0\in \mcc$, which satisfies $a \times 0 = a$ for all $a$ and can usually be omitted from diagrams. Furthermore, each string-net label $a \in \mcc$ has a unique dual label $\overline{a} \in \mcc$ such that $N^0_{a\overline{a}} = N^0_{\overline{a}a} = 1$.

The quantum dimension $d_a$ of a string type $a$ is defined as
%as the factor picked up when removing an $a$-bubble from a diagram:
%\be \includegraphics[scale=.85]{quantum_dimension_def.pdf} \ee
%where $|0\rangle$ denotes the vacuum state. Mathematically, $d_a$ is
the largest left eigenvalue of the fusion coefficient matrix $N_a$ (with entries $[N_a]_{bc} = N_{ab}^c$). Physically, $1/d_a^2$ gives the probability that $a\times \overline{a}$ will fuse to the vacuum. This implies that for Abelian theories, $d_a = 1$ for all $a\in \mcc$. 

\begin{comment}The quantum dimensions form a one-dimensional representation of the fusion algebra:
\be d_ad_b = \sum_c N^c_{ab}d_c. \ee
In particular, the quantum dimensions must satisfy $d_a = d_{\overline{a}}$ for all $a$, and we can always work in a gauge where they are real and positive. 
\end{comment}

The building blocks of string-net diagrams are trivalent vertices, which form a basis of the system's Hilbert space:
\be \label{eq:basis_vertices} \includegraphics[scale=.85]{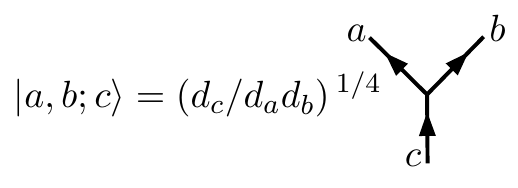},\ee
\be \includegraphics[scale=.85]{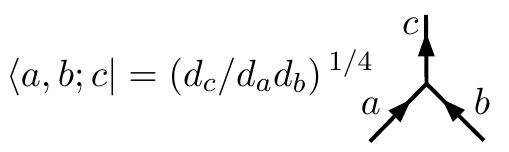}.\ee
The factors involving the quantum dimensions are simply a convenient normalization choice. We call basis vectors of the form $|a,b;c\rangle$ splitting vertices, and dual basis vectors of the form $\langle a,b;c|$ fusion vertices. The dimensions of the spaces $|a,b;c\rangle$ and $\langle a,b;c|$ are given by $N_{ab}^c$, and so a vertex $|a,b;c\rangle$ or $\langle a,b;c|$ will be nonzero only when $N_{ab}^c \neq 0$, in which case we say that the vertex is {\it stable}. More precisely, we say that a vertex $|a,b;c\rangle$ is stable only if the trivial representation appears with non-zero multiplicity in the direct sum decomposition of $a\tp b \tp \bar c$. 

\begin{comment}
Given a splitting vertex $|a,b;c\rangle$ and a fusion vertex $\langle a,b;d|$, we can form their inner product by stacking $\langle a,b;d|$ on top of $|a,b;c\rangle$:
\be \label{eq:inner_product_def} \includegraphics[scale=.85]{inner_product_bubble.pdf} \ee
which reads $\langle a,b;d | a,b;c\rangle = \delta_{cd}\sqrt{d_ad_b/d_c} | c\rangle$. Similarly, their outer product gives a partition of unity through $|a\rangle \tp |b\rangle = \sum_c \sqrt{d_c/d_ad_b} | a,b;c\rangle \langle a,b;c | $, where $|a\rangle$ represents a vertical $a$ worldline. Graphically, this looks like
\be 
\includegraphics[scale=.85]{partition_of_unity.pdf}
\ee
\end{comment}

The most basic deformation of a string-net graph is called the $F$-move, and is given by a unitary transformation which implements a change of basis between the two different ways of decomposing the space $|a,b,c;d\rangle$ into trivalent splitting vertices:
\be \label{eq:F_move} \includegraphics[scale=.85]{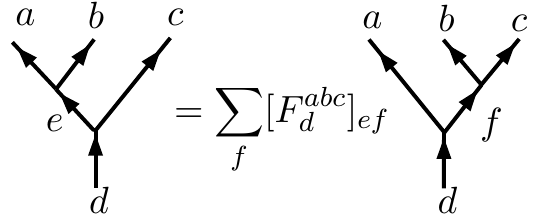} \ee
%Moving the string $b$ from the right to the left of the diagram is done with the inverse of the $F$-move:
%\be \includegraphics[scale=.85]{F_move_inv.pdf} \ee
We will always work in a unitary gauge where all the $F^{abc}_d$ are unitary matrices\cite{Bais14, Lan14}:
\be [F^{abc}_d]^{\dagger}_{fe} = [F^{abc}_d]^*_{ef} = [F^{abc}_d]^{-1}_{fe}. \ee
%In this convention, the quantum dimensions are determined from the $F$-symbols by the gauge invariant quantity
%\be d_a = \frac{1}{\sqrt{[F^{a\bar a a}_{a}]_{00} [F^{\bar a a \bar a}_{\bar a}]_{00}}}. \ee 
The unitarity of the $F$-symbols allows us to normalize the fusion data by setting $[F^{abc}_{d}]_{ef} = 1$ whenever one of $a,b,$ or $c$ is the vacuum label. It is often assumed that the $F$-symbols are invariant under a certain action of the tetrahedral group \cite{Levin05,Lan14,Hu15}. We will not assume tetrahedral symmetry in this paper, since the time-reversal and parity breaking phases we are interested in will necessarily break this tetrahedral symmetry (we derive more general tetrahedral symmetry relations in appendix \ref{sec:tetra}). 
 
A diagrammatic manipulation that will be very important later on is the ability to straighten wiggles of an anyon worldline. We can straighten wiggles at the cost of a phase factor $\alpha$, which depends on the label of the worldline:
\be \label{eq:FS_def}
\includegraphics[scale=.85]{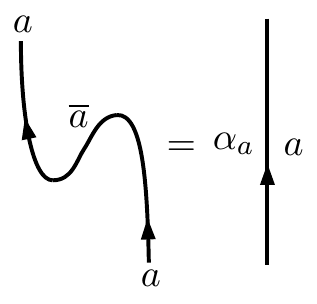}
\ee
The quantity $\alpha_a$ is known as the {\it Frobenius-Schur indicator} of $a$, and is related to the fusion data through $\alpha_a = d_a [F^{a\overline{a}a}_{a}]_{00}$. It is always possible to choose an appropriate gauge\cite{Rowell07,Bais14} in which $\alpha_a = 1$ if $\bar a \neq a$ and $\alpha_a = \pm1$ if $\bar a = a$.%, and our normalization convention in (\ref{eq:basis_vertices}) guarantees that this gauge is automatically chosen.
 
Another key diagrammatic manipulation that we will need to make use of is a way to transform between splitting and fusion vertices. By using the $F$-moves and our ability to add / remove wiggles using (\ref{eq:FS_def}), it is easy to show that (see appendix \ref{sec:more_diagrams} for details):
\be \label{eq:lowering_operators} \ba & \includegraphics[scale=.85]{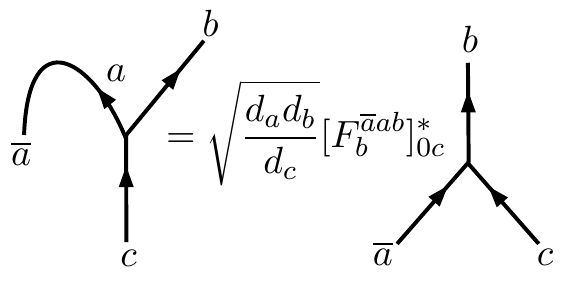} \\ & \includegraphics[scale=.85]{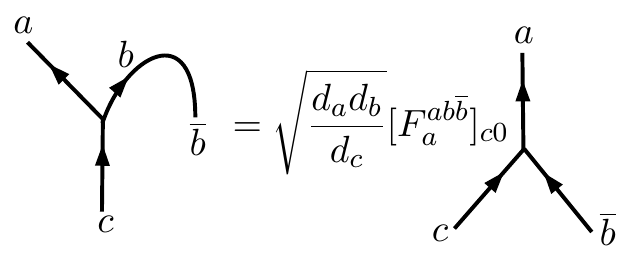} \ea \ee 
To transform a fusion vertex to a splitting vertex in a similar fashion, we simply conjugate the relevant operator. 

In addition to sliding anyon worldlines around and manipulating the order in which they fuse, we would like to be able to describe processes which braid anyons around one another. To this end, we define a device for quantifying the twisting of two worldlines around one another, known as the $R$ matrix:
\be \includegraphics[scale=.85]{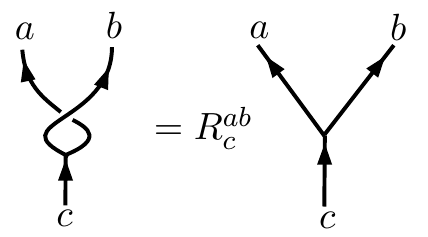}. \ee
The topological twist $\theta_a$ of an anyon $a$ is defined by the phase factor picked up when untwisting a loop in a worldline labeled by $a$:
\be \label{eq:twist_def} \includegraphics[scale=.85]{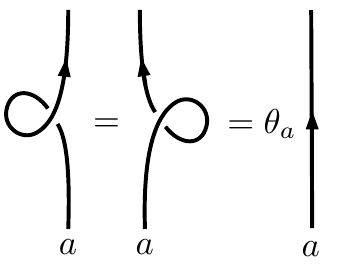}. \ee
The twist $\theta_a$ can also be expressed by averaging over the entries of the $R$-matrix $R^{aa}$ weighted by the quantum dimensions:
\be \theta_a = \sum_{b\in \mcc} \frac{d_b}{d_a} R^{aa}_b. \ee
Finally, we point out that braidings like
\be \includegraphics[scale=.85]{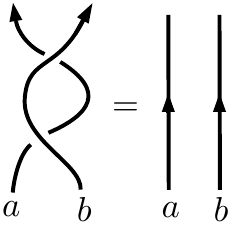} \ee
are always trivial, and can as such can be freely added or removed from diagrams.

A theory is said to be {\it braided} if the twisting action of the $R$ matrix commutes with the $F$-moves, in a sense which is made precise in appendix \ref{sec:braiding}. Most work on string-net models has focused on braided theories, although we will see that slightly relaxing the braiding condition can lead to several physically interesting consequences like broken time-reversal and parity symmetry. Indeed, considering non-braided theories will be necessary for realizing the type of rotational symmetry fractionalization considered in this paper. 

All of the diagrammatic machinery described above is written in terms of trivalent vertices, although in general we would like to consider string-net graphs defined on graphs of arbitrary valence. Fusion and vertex stability are still well-defined for general graphs: if $\{ a_i\}$ label all the outgoing worldlines of a vertex and $\{b_i\}$ label all the incoming lines, there is certainly no obstacle to formally taking the fusion product $\prod_ia_i \times  \prod_j \overline{b_j}$ and constructing more general fusion coefficients $N_{a_1\dots a_n}^{b_1\dots b_m}$. In analogy with the trivalent case, the vertex $|a_1,\dots,a_n;b_1,\dots,b_m\rangle$ will be stable (i.e. nonzero) only when $N_{a_1\dots a_n}^{b_1\dots b_m}\neq 0$. We will see that working with more general graphs allows for a more natural description of the invariant computed in this paper, which involves the fractionalization of a certain form of rotational symmetry over collective groups of multiple (e.g., more than three) anyons, and which is most easily described by working with graphs of arbitrary valence. 

Finally, we should mention that throughout this paper we will focus only on topological phases defined on {\it closed} manifolds with trivial topology. Relaxing the latter constraint is straightforward, although relaxing the former entails a careful treatment of boundary conditions, which we leave to future work. 

\section{A form of symmetry fractionalization in the tensor product state construction}\label{sec:TPS}

A natural motivation for the invariant we consider in this paper comes from the tensor product state construction, which is a sort of mean-field description of states with topological order\cite{Verstraete08,Gu09b,Buerschaper09}. Tensor product states are fixed-points under a natural wave function renomalization group transformation and efficiently encode the long-range entanglement of topologically ordered states \cite{Vidal07,Aguado08}. Furthermore, an explicit wave function for the ground state of string-net phases\cite{Gu09,Buerschaper09} can be constructed as a tensor product state, by performing a weighted tensor trace over the system's Hilbert space.

We construct the tensor product state wave function for our string-net models as follows, by slightly generalizing the procedure in Ref.\cite{Gu09}. For each $n$-valent vertex $|a_1,\dots,a_n\rangle$ in the string-net graph, we form an $n$-index tensor $T^{a_1\dots a_n}$. Here, each $a_i$ is an index which transforms according to the representation given by the worldline $a_i$. We let $a_i$ be an upper (lower) index of $T$ if the associated worldline labeled by $a_i$ points out of (into) the vertex. Furthermore, we set $T^{a_1\dots a_n} = 0$ if $|a_1,\dots,a_n\rangle$ is not stable, i.e. if the vacuum string $0$ does not occur in the fusion product $a_1 \times \dots \times a_n$.

To compute the string-net wavefunction, we construct one such $T$-tensor for each vertex of the string-net graph and tensor all such $T$-tensors together to form a single giant tensor. The wave function is then calculated by performing a weighted contraction (known as a weighted tensor trace) over the repeated indices of this tensor, which are the physical degrees of freedom indexed by the labels of the anyon worldlines. That is, 
\be \Psi_{\text{string-net}} = \sum_{\big\{\substack{\text{worldline} \\ \text{~labels } a_i}\big\}} \text{wtTr}\bigg[ \bigotimes_{\big\{\substack{ {\rm vertices}\\ |a_1,\dots,a_n\rangle}\big\}} T^{a_1\dots a_n} \bigg]. \ee
The weight in the weighted tensor trace consists of products of the quantum dimensions in the theory\cite{Gu09}, the details of which are not important for us. The wavefunction $\Psi_{\text{string-net}}$ is a number as it must be, since each $a_i$ appears in $\bigotimes_{{\rm vertices}} T^{a_1\dots a_n}$ once as an upper index and once as a lower index. 

We now consider what happens when we have a $T$-tensor whose indices are all either upper or lower and all transform under the same representation. A priori, this tensor may not be cyclically symmetric, i.e. we may have $T^{a_1\dots a_n} = \mco^{(v)}_n T^{a_n\dots a_{n-1}}$ for some factor $\mco^{(v)}_n$ which depends on the labels of the vertex $v = |a_1,\dots,a_n\rangle$. This implies an ambiguity in our construction for the $T$ tensors and thus for the final string-net wave function, as was pointed out in Ref.\cite{Lin14} Somewhat surprisingly, $\mco^{(v)}_n$ is {\it not} always a gauge degree of freedom -- indeed, $\mco^{(v)}_n$ is the invariant mentioned in the introduction that we will employ to test the parity and time-reversal invariance of the topological phase described by $\mcc$.

To remedy this ambiguity, we build on the construction in Ref.\cite{Lin14} and assign a branch cut to each vertex that tells us how to assign the legs of the vertex to the indices of its associated $T$-tensor. We choose the convention that the indices of $T$ are indexed counterclockwise from the position of the branch cut. For example, for a tetravalent vertex $|a,b,c,d\rangle$ we have 
\be \includegraphics[scale=.85]{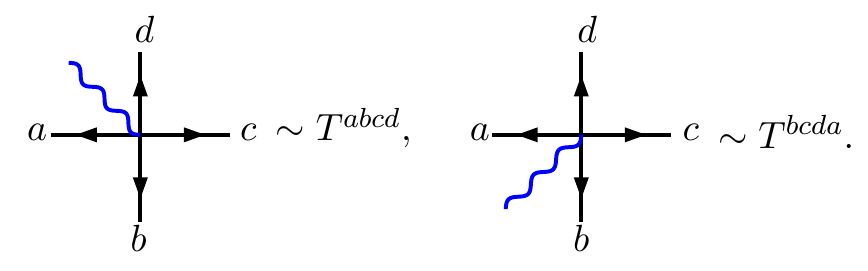} \ee
where the blue wiggly line denotes the branch cut. By introducing the branch cuts into our graphical calculus, we generalize the dots of Ref. \cite{Lin14} to more general types of lattices. Our extension is also simpler than the devices introduced in Ref.\cite{Lin14}, since in our perscription we don't have to keep track of vacuum strings and their orientation.

Although we find it helpful to include vertices of arbitrary valence into our generalized string-net graphs to group bunches of anyons together, all of the diagrammatic machinery developed in Section \ref{sec:review} was written in terms of trivalent vertices. To be able to do computations with these more general string-net graphs, we decompose each $n$-valent vertex into a collection of trivalent vertices and then perform a tensor contraction over the internal degrees of freedom at each vertex created during the decomposition. %Since the three-index $T$-tensors can be written in terms of the fusion data\cite{Gu09}, our resulting $n$-index tensors can also be written in terms of the fusion data. 
%Even though we will usually perform calculations with trivalent graphs, we will see that considering vertices of higher valence will be helpful in realizing certain emergent properties that different collections of anyons possess. 

Of course, there are many possible ways of decomposing an $n$-valent vertex into a network of connected trivalent vertices. All decomposition choices are equivalent to one another due to the unitarity of the $F$-symbols (or more precisely, due to the coherence theorem of category theory\cite{Maclane78}), and so we choose an internal structure that will facilitate computations made later on. A generic vertex in the string-net graph will possess wordlines oriented both into and out of the vertex, although we observe that by repeatedly performing unitary transformations on the vertex using the local rules (\ref{eq:lowering_operators}) and (\ref{eq:FS_def}), we can transform the trivalent decomposition of any $n$-valent vertex into one where in a local region around the vertex all of its external worldlines are oriented either all outwards or all inwards, possibly after the creation of additional 2-valent vertices outside of the immediate vicinity of the vertex. Furthermore, we can also ensure that all of the vertex's internal vertices are either all splitting (if the vertex's external worldlines are outgoing) or all fusion (if they are incoming). 

For example, we can decompose the tetravalent vertex $|b,c,d;a\rangle$ into a structure with all outgoing worldlines and only splitting internal vertices as 
\be \includegraphics[scale=.85]{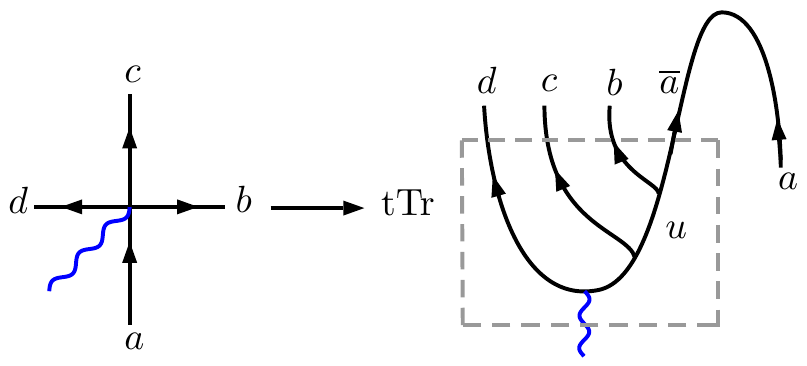} \ee
where tTr indicates a summation over the internal degree of freedom $u$, and the dashed box encloses the original decomposed tetravalent vertex. We use the dashed box to denote how the branch cut organizes the outgoing legs of the decomposed vertex -- the indices of the $T$-tensor for the decomposed vertex are labeled according to the worldlines which leave the dashed box, proceeding counterclockwise from the branch cut. Worldlines completely contained within the dashed box (like the worldline $u$ in the above example) always need to be traced out. This means that the region enclosed by the dashed box in the above diagram would be assigned the tensor $T^{\overline{a}bcd}$. In the example above, we have thus transformed a vertex with one incoming worldline into a vertex with four outgoing worldlines and an extra 2-valent vertex. In terms of the $T$-tensors, this decomposition reads 
\be T_a^{bcd} \rightarrow \ttr( T_{a\overline{a}} \tp  T^{\overline{a}b}_u \tp T_{\overline{d}}^{uc}\tp T^{\overline{d}d} ).\ee

After performing operations like this at every vertex of the string-net graph, we will be left with a lattice where each vertex possesses either only outgoing or only incoming worldlines. For example, if the string-net model is derived from a square lattice, we might perform the decomposition as 
\be \includegraphics[scale=.85]{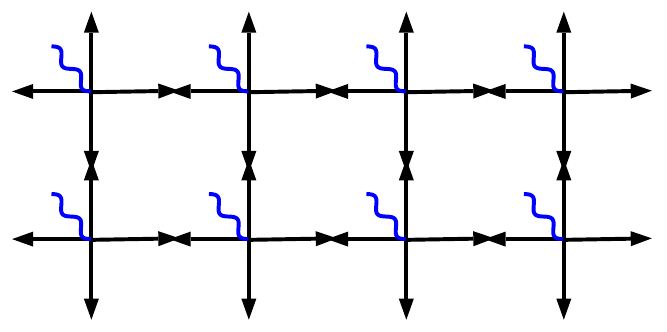}, \ee
where we have omitted branch cuts at each of the 2-valent vertices for simplicity. 

In the discussion that follows, we will focus on vertices with only outgoing wordlines and whose internal structures are entirely composed of splitting vertices. An identical analysis can be carried out for vertices with only incoming worldlines and fusion vertices with trivial modifications by using the unitarity of the $F$-symbols. By focusing on vertices with only outgoing worldlines, we are allowed to leave the arrows on the anyon worldlines (which will always be pointing upwards) implicit in local diagrams, which we will do from now on. In this convention, the flow of time is always directed upwards in diagrams. We stress that our convention of orienting all outward-pointing worldlines upwards in diagrams is nothing more than a convenient notational choice that will make calculations simpler later on, allowing us to build in the time direction into the structure of the diagrams rather than keeping track of it by placing arrows on every worldline. %With this convention, the flow of time always points upwards in diagrams. 

This ``upward flow of time'' notation does not by itself break the {\it spatial} rotational invariance of the original string-net graph -- it is merely a convenient notational choice. In this notation, 
rotating the graphs in the plane of the paper will not generically leave them invariant, since this involves doing a rotation in time as well as in space. However, putting aside the branch cuts, they are still invariant under spatial rotations: we may simply restore the arrows on every worldine and rotate the diagram in the plane of the paper while keeping the directions of the arrows fixed, which leaves the physics of the string-net graph unchanged. Thus, the effects of performing a spatial rotation on the string-net graph are captured entirely by the branch cuts, which we orient in a given direction so that they remain fixed in place during a spatial rotation. This means that performing a physical rotation of the string-net graph (which exchanges different lattice sites in a given lattice realization) is equivalent to leaving the string-net graph fixed, but performing a rotation of the branch cuts at each vertex, which is an {\it onsite} operation.

%Choosing a preferred ``up'' direction for our diagrams breaks the rotational symmetry of the diagrams, but in this convention a rotation of the graph is actually a rotation in time as well, which may require reversing the directions of the arrows on the worldlines. This is just saying that graphs may not be symmetric upon reversing the flow of time. If we re-draw the arrows on the worldlines to explicitly keep track of the flow of time for each individual worldline, the rotational invariance of the diagrams themselves is restored. 

The upward flow of time notation also allows us to fix a convenient decomposition pattern for each $n$-valent vertex into a collection of splitting vertices. The decomposition we choose is written graphically as  
\be \label{eq:internal_Ttensor_def} \includegraphics[scale=.85]{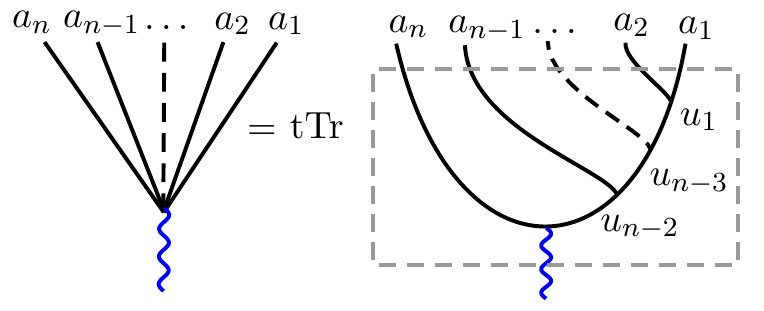}. \ee
Here, the $u_i$ label the internal degrees of freedom at each vertex created during the decomposition into trivalent vertices, which need to be traced out. As before, we form the indices of the vertex's $T$-tensor by reading off the labels of the worldlines leaving the dashed box proceeding counterclockwise from the branch cut (giving $T^{a_1\dots a_n}$ for the above diagram). Additionally, in the above diagram we've used the dashed line to represent a collection of similar splitting vertices:
\be \label{eq:dashed_line_clarification} \includegraphics[scale=.85]{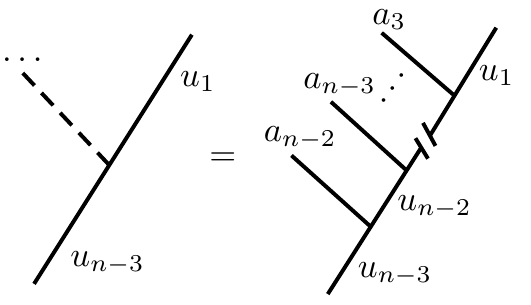}\ee
Given a vertex $|a_1,\dots,a_n\rangle$, we can thus calculate the associated tensor $T^{a_1\dots a_n}$ by tensoring together the 3-index $T$-tensors associated with each internal vertex and then performing a tensor trace over the internal degrees of freedom $\{u_i\}$ (note that each $u_i$ will appear once as a lower index and once as an upper index). This gives
\be %explicitly 
%T^{a_1 \dots a_n} =  \ttr \left[T^{a_1a_2}_{u_1} \otimes T^{a_3 u_1}_{u_2} \otimes \dots \otimes T^{a_{n-1} u_{n-3}}_{u_{n-3}}\tp T^{a_n u_{n-2}}_0 \right] 
%or as a product
T^{a_1\dots a_n} = \ttr \left( \bigotimes_{k=1}^{n-1} T^{u_{k-1}a_{k+1}}_{u_k} \right) 
\ee
with $u_0 = a_1$ and $u_{n-1} = 0$. 

At this point, we should note that we have made no mention of a Hamiltonian for our generalized string-net construction. It turns out that slightly modifying the standard string-net Hamiltonian\cite{Lin14,Levin05} by accounting for the branch cuts and arbitrary lattice structure is sufficient for our purposes. Since this is not the main focus of our paper, we relegate a discussion of the Hamiltonian to appendix \ref{sec:Hamiltonian}. 

As mentioned earlier, the $T$-tensors may not be cyclically symmetric, and so the way in which they transform under cyclic permutations of their indices induces a $\ZZ_n$ action on the vertices of the graph. 
There are two ways to interpret this $\ZZ_n$ action. One way is to think of it as an internal symmetry, with an application of the $\ZZ_n$ symmetry action manifesting itself as a rotation of the internal branch cuts on each vertex counterclockwise by an angle $2 \pi /n$ so that they each pass over one anyon worldline. Alternatively, we can choose a globally defined preferred orientation for our string-net graphs (as in refs. \cite{Lan14,Kong12}), and fix the orientation of the branch cuts along this direction (as is usually done in trivalent string-net models by evaluating all vertices clockwise or counter-clockwise starting from the `top' of the vertex). In this interpretation, a global rotation of the branch cuts at every vertex by an angle $\theta$ is equivalent to a physical rotation of the underlying string-net condensate by $\theta$ with respect to this preferred direction. 
We can thus express the action of a global symmetry operator $\hat\mco_{n}$ which physically rotates the underlying string-net graph $|SN\rangle$ by an angle $2\pi/n$ as a product 
\be \hat\mco_{n}|SN\rangle = \prod_{\{\text{vertices~}v\}} \hat\mco_{n}^{(v)} |SN\rangle, \ee
where each $\hat\mco_{n}^{(v)}$ is a local operator that is supported on a region localized around the $v^{th}$ vertex in the product and which performs a rotation of the $v^{th}$ vertex's branch cut so that the indices of the $T$-tensor associated with the vertex $v$ are cyclically permuted. We emphasize that the $\hat\mco_{n}^{(v)}$ operators only implement {\em orientation-preserving} rotations, corresponding to cyclic permutations of the $T$-tensor indices. We do not consider {\em orientation-reversing} reflections of the string-net graph. Additionally, we note that the form of the $\mco_n^{(v)}$ operators implies that they commute with our generalized string-net Hamiltonian (see Appendix \ref{sec:Hamiltonian}).

The action of the operators $\hat\mco^{(v)}_n(T^{a_1\dots a_n}) = \mco_n^{(v)} T^{a_n\dots a_{n-1}}$ which implement the branch cut rotation are computed at each vertex by moving the vertex's branch cut counterclockwise by $2\pi/n$. Equivalently, we can compute $\mco^{(v)}_n$ by deforming the internal structure of the vertex $v$ so that the order of its outgoing worldlines with respect to its branch cut is cyclically permuted. Diagrammatically, this is illustrated by the process
\be\label{eq:def_of_onsite} \includegraphics[scale=.85]{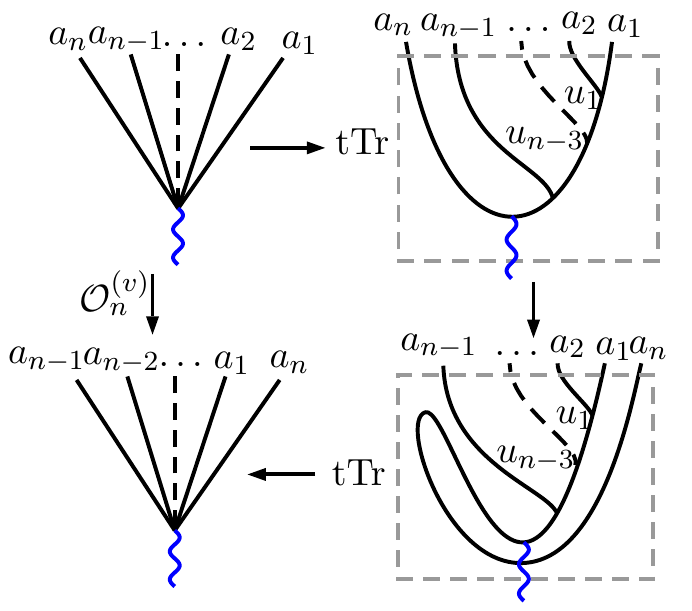}. \ee

Since rotating the branch cut on an $n$-valent vertex by an angle $2\pi/n$ $n$ times is the identity operator, we see that the $\hat\mco_n^{(v)}$ operators form a $\ZZ_n$ group structure described by the relation $[\hat\mco_n^{(v)}]^n|a_1,\dots,a_n\rangle =|a_1,\dots,a_n\rangle$. Thus, the vertices of the string-net graph, which represent physical multi-anyon states, transform linearly under the branch cut rotation. To explore the possible fractionalization of this symmetry, we must examine the action of $\mco_n^{(v)}$ on the single-anyon states, which are the individual anyon worldlines $\{|a\rangle\}$.  

Suppose that $|a_i\rangle$ is a worldline emanating from the vertex $v=|a_1,\dots,a_n\rangle$. When $\hat\mco_n^{(v)}$ acts on the vertex $v$, it will only act nontrivially on the outgoing worldline of $v$ which the branch cut is passed over during the rotation. 
This means that $\hat\mco_n^{(v)}|a_i\rangle = \mco_n^{(v)}|a_i\rangle$ if the branch cut passes over the worldline $|a_i\rangle$ during the implementation of $\hat\mco_n^{(v)}$, while $\hat\mco_n^{(v)}|a_i\rangle = |a_i\rangle$ if it does not. This implies that if we apply $\hat\mco_n^{(v)}$ $n$ times to the vertex $v$, only one of the $\hat\mco_n^{(v)}$ factors will act nontrivially on a given outgoing worldline of $v$. We then see that the action of the branch cut rotation on the single-anyon state $|a_i\rangle$ is given by $[\hat\mco_n^{(v)}]^n|a_i\rangle = \mco_n^{(v)}|a_i\rangle$. Since $\mco_n^{(v)}$ can generically be any $n^{th}$ root of unity, this means that the individual anyon worldlines transform {\it projectively} under rotation, and so the single-anyon states indeed carry a ``fractionalized'' rotational quantum number. 

Before moving on, we make two adjustments to our notation. Firstly, from now on we will leave the dashed boxes implicit in the diagrams. Secondly, since $\mco_n^{(v)}$ is only nontrivial when the vertex $v$ is of the form $|a\rangle^{\tp n}$, we will define $\mco_n^{(v)} =1$ when $v$ is not of the form $|a\rangle^{\tp n}$. For vertices of this form, we will switch to the more transparent notation $\mco_n(a)$.

\section{Computing the $\mco_n$ indicators} \label{sec:onsite}

\subsection{The $\mco_n$ indicators and time-reversal and parity symmetry}

Our main motivation for studying the $\mco_n$ indicators lies in the following theorem:
\begin{customthm}{1}\label{mcotheorem}
If $\mco_n(a)$ is not real for at least one choice of $n\in \ZZ$ and $a\in\mcc$, then the topological phase described by $\mcc$ must break parity and time-reversal symmetry. 
\end{customthm}
We present a proof of this theorem in Appendix $C$. 

The above theorem is useful since the indicators $\mco_n$ are often quite straightforward to compute, as we will demonstrate in the following section. In particular, we will derive an expression for the $\mco_n$ that {\it only} involves the fusion data $\{[F^{abc}_d]_{ef},N^{c}_{ab},d_a\}$. This means that computing the invariants $\mco_n(a)$ can provide us with an efficient way to test whether or not a given topological phase is time-reversal or parity symmetric by using only the $F$-symbols and fusion rules of the theory. This frees us from testing time-reversal and parity symmetry compatibility by looking at braiding-related information like the chiral central charge or modular $S$ matrix, the computation of which is often fairly arduous\cite{Lan14,Bonderson07}. On the other hand, we also derive an expression for the $\mco_n$ that involves only the topological twists and fusion rules $\{\theta_a,N^c_{ab}\}$ (although for non-Abelian theories, more braiding information may be needed). This provides a test of time-reversal and parity invariance without using solving the pentagon equations to obtain the $F$-symbols, which is computationally more efficient if one starts with $\{\theta_a,N^c_{ab}\}$ as the defining data of the theory (as in ref\cite{lan2015theory}). 
%Additionally, we believe that this second formula for the indicators $\mco_n$ can be applied even to theories with nontrivial chiral central charge which do not admit a traditional string-net description.

In the following sections, we derive these two ways of computing the $\mco_n$ indicators. We will start with the expression for the $\mco_n$ in terms of the fusion data $\{[F^{abc}_d]_{ef},N^{c}_{ab},d_a\}$. 

\subsection{The $\mco_n$ indicators in terms of the $F$-symbols} \label{sec:fusiondata_mco}

%We proceed to determine explicit expressions for the phases $\mco_n$ corresponding to the $\ZZ_n$ rotational symmetry fractionalization. 
%This provides a generalization of the $\alpha$ and $\gamma$ factors introduced in Ref.\cite{Lin14} by extending their results to non-Abelian phases and extending the branch cuts to more general anyon structures (beyond just groups of three anyons). 

As a warm-up, we compute the $\mco_2$ rotational symmetry action for 2-valent vertices, which we will interpret as trivalent splitting vertices with the incoming worldline labeled by the vacuum. In order for the vertex to be stable, it must be of the form $|a,\overline{a};0\rangle$. Since the $\ZZ_2$ symmetry action will be nontrivial only when both indices of $T^{a \overline{a}}$ transform under the same representation, we see that the $\ZZ_2$ symmetry action is only nontrivial when $a = \overline{a}$. With the help of (\ref{eq:FS_def}) to insert a wiggle in the diagram, we compute the symmetry action as 
\be \includegraphics[scale=.85]{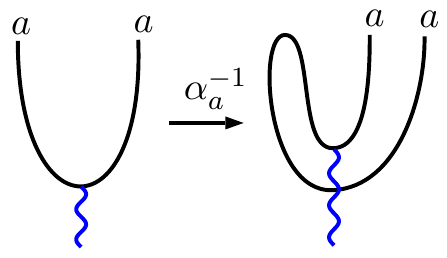}. \ee
Since the physical two-particle state $|a,a;0\rangle$ must transform trivially under a complete rotation of the branch cut, the individual single-particle states $|a\rangle$ can transform projectively by a $\ZZ_2$ phase. That is, we must have $\alpha_a = \alpha_{\overline{a}}\in\ZZ_2$. 
Thus, we have shown that the rotational symmetry action for two-valent vertices is identical to the Frobenius-Schur indicator, i.e. $\mco_2(a) = \alpha_a = \pm 1$ ($\mco_2$ is also the same as the $\gamma$ factor introduced in ref.\cite{Lin14}). We will now study the fractionalization of a collection of $n \geq 3$ identical anyons by computing $\mco_n$ for general $n$, which correspond to the invariants in category theory known as higher Frobenius-Schur indicators \cite{Ng07,Kashina06}. 

%%THE GENERAL SUM OVER STUFF FORMULA FOR NON-ABELIAN GUYS

As in the previous section, we decompose each $n$-valent vertex according to the prescription in (\ref{eq:internal_Ttensor_def}). The rotational symmetry action corresponds the process of moving the leftmost outgoing worldline of the vertex around to the right side of the vertex so that it passes through the vertex's branch cut, as in (\ref{eq:def_of_onsite}). As mentioned previously, the rotational symmetry action on a given vertex will be nontrivial only when all the labels of the vertex's outgoing worldlines are identical. So then assuming that $a$ is an anyon label such that the decomposition of $a^{\times n}$ contains the trivial representation with non-zero multiplicity (i.e., assuming that the vertex $|a\rangle^{\tp n}$ is stable), we calculate the rotational symmetry action abstractly through the following diagram:
\be \label{eq:onsite_F_diagrams} \includegraphics[scale=.9]{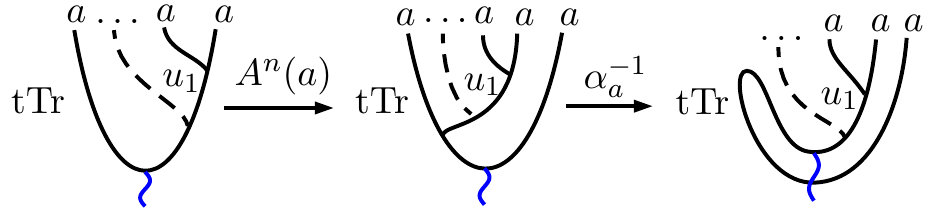},\ee
where the dashed line is to be understood as in (\ref{eq:dashed_line_clarification}), but with all outgoing legs labeled by $a$. If the vertex $|a\rangle^{\tp n}$ is not stable, we set $\mco_n(a) = 0$. 

The map $A^n(a)$ is the map which moves all the legs branching off on the right hand side of the vertex structure to the left side of the vertex structure through a combination of $F$-moves, as defined by the first arrow in (\ref{eq:onsite_F_diagrams}). $A^n(a)$ can thus be thought of as a ``big inverse $F$-move'' applied to the entire internal structure of the vertex. To proceed, we need to write $A^n(a)$ explicitly in terms of the fusion data. To do this, we derive the action of $A^n(a)$ by applying inverse $F$-moves to the vertex, starting from the top right of the diagram and proceeding down until all the worldlines have been moved to the left hand side. This process creates additional internal degrees of freedom $\{v_i\}$, which must be traced out. For example, the first few steps in the computation of $A^n(a)$ are:
\be \includegraphics[scale=.85]{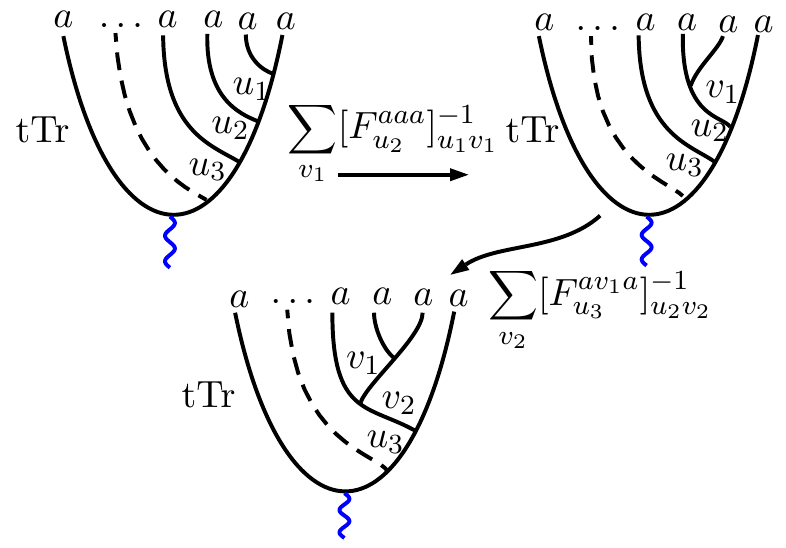} \ee
where the tTr's remind us that the internal degrees of freedom $\{u_i\}$ are being summed over. Proceeding inductively, we see that 
\be \label{eq:mco_F_symbols} \ba A^n(a) = \sum_{v _1\dots v_{n-2}} [F^{aaa}_{u_2}]^{-1}_{u_1v_1} [F^{av_1 a}_{u_3}]^{-1}_{u_2v_2} \dots [F^{a v_{n-3} a}_0]^{-1}_{u_{n-2}v_{n-2}}. \ea \ee
Using the definition of the regular Frobenius-Schur indicator ($\mco_2$) and the fact that $\alpha_a^{-1} = \alpha_a $ for any $a\in \mcc$, we insert a wiggle in the diagram in order to proceed along the bottom right arrow in (\ref{eq:def_of_onsite}). We then notice that in order to match the definition in (\ref{eq:def_of_onsite}), the internal vertices $\{v_i\}$ must match the initial labels $\{u_i\}$, since we want the action of the operator $\hat\mco_n$ to preserve the internal structure of the diagram while moving its leftmost outgoing worldline around the branch cut. This means that we are required to project $v_i = u_i$ for all $i$, and so
\be \ba \label{eq:onsite_F_equation}\mco_n(a) = \alpha_a \sum_{\{u_i\}} \prod_{k=1}^{n-2} [F^{au_{k-1}a}_{u_{k+1}}]^{-1}_{u_ku_k}, \ea \ee
where we can set $u_0 = a$, $u_{n-1} = 0$, and $u_{n-2} = \overline{a}$. 
%Note that the unitarity of the $F$-symbols ensures that $\mco_2(a) = \alpha_a$ if $|a,a;0\rangle$ is stable, and $0$ otherwise.  

This expression can be greatly simplified if the fusion rules are derived from the group structure of a finite group $G$. In this case, the anyon labels in $\mcc$ are associated with the elements of $G$, and anyon fusion is given simply by group multiplication in $G$. This implies that the knowledge of any two labels on the legs of a non-zero trivalent vertex uniquely determines the label of the third leg, which greatly simplifies the computation of $A^n(a)$, since a stable vertex $|a\rangle^{\tp n}$ must decompose into trivalent vertices as 
\be \label{eq:grading_decomp} \includegraphics[scale=.85]{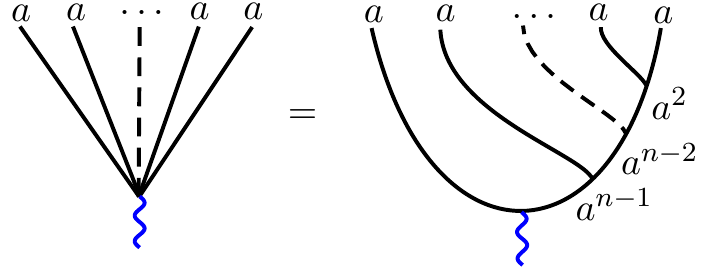} \ee
where we have omitted the sum over the internal degrees of freedom since it simply projects onto the configuration of worldline labels shown above. Using this decomposition, we see that the product in (\ref{eq:onsite_F_equation}) is only non-zero when $u_k = a^{k-1}$, and so we can write
\be \mco_n(a) = \delta_{a^n,1}  \prod_{k=1}^{n-1} [F^{aa^{k-2}a}_{a^k}]^{-1}_{a^{k-1}a^{k-1}} \ee
where the $\delta$ function enforces the requirement that the vertex $|a\rangle^{\tp n}$ is stable. 

For theories derived from the group structure of a finite group $G$, the non-zero $F$-symbols take the form \cite{Propitius95}
\be [F^{a,b,c}_{abc}]_{ab,bc} = \omega(a,b,c), \ee
where $\omega(a,b,c)$ is a function taking values in ${\rm U}(1)$ and satisfying the algebraic relation 
\be \omega(a,b,c)\omega(a,bc,d)\omega(b,c,d) = \omega(ab,c,d)\omega(a,b,cd),\ee
which follows directly from the pentagon identity. Such functions are known as 3-cocycles, and topologically distinct classes of such functions are parametrized by the third cohomology group $H^3(G,{\rm U}(1))$. It then follows from (\ref{eq:onsite_F_equation}) that 
\be \label{eq:onsite_F_equation_finiteG} \mco_n(a) = \delta_{a^n,1} \prod_{k=1}^{n-1} \omega^{-1}(a,a^k,a) \ee
This formula is very helpful for computations involving phases derived from finite groups or simple SPT and SET phases, and we will use it extensively in Section \ref{sec:examples}. 

We note that since these phases are parametrized by $H^3(G,{\rm U}(1))$, cohomologically equivalent $\omega$ should represent physically identical phases. That is, we should be free to modify $\omega$ by the coboundary gauge transformation 
\be \label{eq:omega_cocycle_gauge} \omega(g,h,k) \mapsto \frac{f(g,hk)f(h,k)}{f(g,h)f(gh,k)} \omega(g,h,k) \ee
where $f : G^2 \ra {\rm U}(1)$ without changing any physical aspects of the theory. Indeed, it is easy to see that the RHS of (\ref{eq:onsite_F_equation_finiteG}) is invariant under equation (\ref{eq:omega_cocycle_gauge}). Since $\mco_n(a)$ does not depend on the choice of representative of the cohomology class of $\omega$, it must also be a class function, by virtue of the easily shown fact that $\omega(mgm^{-1},mhm^{-1},mkm^{-1})$ is cohomologous to $\omega(g,h,k)$ for all $g,h,k,m\in G$. 

In more general scenarios, we can consider altering the $F$-symbols by the gauge transformation 
\be \label{eq:vertex_gauge_tform} [F^{abc}_d]_{ef} \mapsto \frac{z^{ab}_dz^{dc}_e}{z^{bc}_fz^{af}_d}[F^{abc}_d]_{ef} \ee
for some functions $z^{ab}_c$. The invariance of $\mco_n$ under these more general vertex-based gauge transformations will be easily derived after making use of the results obtained in the following section. 

\subsection{The $\mco_n$ indicators in terms of the topological twists}

We will now derive a different way of computing the $\mco_n(a)$ indicators, this time in terms of the topological twists $\{\theta_a\}$. %We believe that, at least for theories defined on closed manifolds, the resulting expressions for the $\mco_n(a)$ apply even to topological phases with nonzero chiral central charge, which do not admit a typical string-net description. 

To compute the rotational symmetry action, we need to move the leftmost outgoing worldline of the vertex to the right side of the vertex's branch cut, as shown in (\ref{eq:def_of_onsite}). The procedure is slightly different for Abelian and non-Abelian theories, and we will start with the simpler Abelian case. Our computation proceeds in three steps, according to the following diagram: 
\be \label{eq:onsite_twist_relation} \includegraphics[scale=.85]{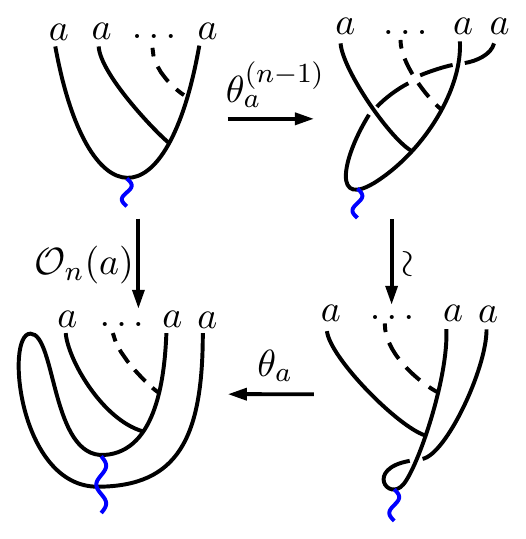},\ee
where the $\xrightarrow{\sim}$ means an equality. 
First, we first drag the leftmost outgoing worldline of the vertex under the other $n-1$ outgoing worldlines. Each time we do this we pick up a factor of $\theta_a$, since
\be \label{eq:abelian_twist} \includegraphics[scale=.85]{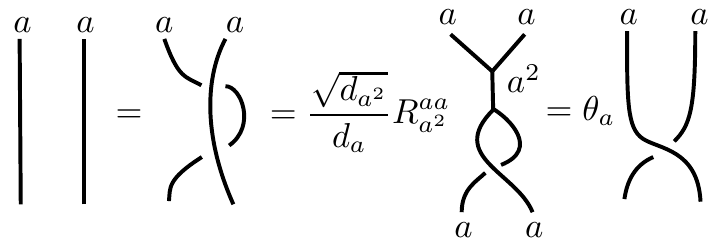} \ee
where we have used the partition of unity 
\be \label{eq:partition_of_unity} |a\rangle \tp |b\rangle = \sum_c \sqrt{\frac{d_c}{d_ad_b}} | a,b;c\rangle \langle a,b;c |.\ee 

Next, we pull this worldline down through the internal structure of the vertex while keeping its endpoints fixed, so that at the end of the movement it passes under only the vertex's bottommost worldline. For theories with an anyonic symmetry that fractionalizes over the anyon worldlines in a nontrivial way, additional phase factors must be added in some cases during this step (see the discussion section). After performing the above steps we are left with a loop in the bottom of the diagram, which we untwist using the relation
\be \includegraphics[scale=.85]{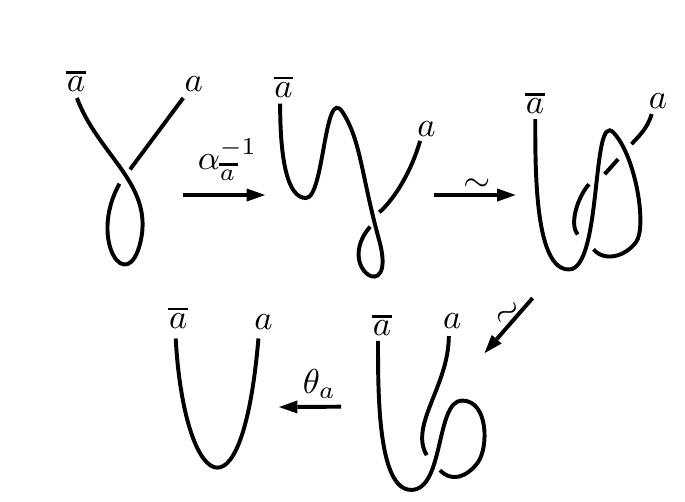}. \ee
Finally, we use (\ref{eq:FS_def}) to insert a wiggle and make the resulting diagram match the one on the right hand side of (\ref{eq:def_of_onsite}). This process is summarized by equation (\ref{eq:onsite_twist_relation}).

%Note that we do not have to worry about the action of the outgoing legs of the vertex on the leg being moved under the vertex during the first step of the computation (\ref{eq:onsite_twist_relation}), since $a_g$ is invariant under $T_g$ (\ref{eq:self_action_invariance}). 
Multiplying all the collected phase factors together, we obtain our final expression for the action of the rotational symmetry. For a collection of $a$ anyons such that the vertex $|a\rangle^{\tp n}$ is stable, we arrive at the simple expression
\be \label{eq:onsite_twist_equation} \mco_n(a) = \theta_{a}^n.  \ee
As before, we let $\mco_n(a) = 0$ if $|a\rangle^{\tp n}$ is unstable. %We note that the derivation of this formula did not depend on the labels of the internal degrees of freedom $\{u_i\}$ at the vertex, and as such is independent of the way in which we decompose the vertex into trivalent vertices. 
Since the topological twists are invariant under gauge transformations on the $F$-symbols, this result serves as a quick but nontrivial check of the invariance of our earlier expression for $\mco_n$  (equation \ref{eq:mco_F_symbols}) under the vertex-based gauge transformation of the $F$-symbols (equation \ref{eq:vertex_gauge_tform}).
%Finally, we should mention that during the three-step computation of $\mco_n$, we could just as well have dragged the $m_g$ worldline {\em over} the rest of the vertex in our calculation of $\mco_n$ (equation \ref{eq:onsite_twist_relation}), rather than under. In this case, the factors of $\theta_{a}$ would become factors of $\theta^{-1}_{a}$, while the product of $U$ factors would become a more complicated product of factors corresponding to the onsite $G$-action on each of the diagram's internal vertices (which correspond to the $U$ factors in ref.\cite{Barkeshli14}). However, this more complicated formula would give the same expression for $\mco_n$, since the relation (\ref{eq:onsite_F_equation_finiteG}) for $\mco_n$ does not contain any mention of braiding-related phases. 

Equation (\ref{eq:onsite_twist_equation}) for $\mco_n$ has several disadvantages. If the phase $\mcc$ includes symmetry defects, then the derivation of (\ref{eq:onsite_twist_equation}) involves braiding symmetry fluxes around one another. While this process is perfectly well-defined mathematically, it is on shaky ground from a physical perspective, since the symmetry fluxes are semi-classical objects and as such are generally not free to be moved around at will. More importantly, the derivation of (\ref{eq:onsite_twist_equation}) does not quite work for non-Abelian theories, since equation (\ref{eq:abelian_twist}) does not hold when there are multiple fusion channels for the product $a \times a$.

To get around these issues, we study the phase consisting of the deconfined quasiparticle excitations of $\mcc$, known as the {\it Drinfeld center} of $\mcc$, written as $\mcz(\mcc)$\cite{Kassel12,Etingof05,Gelaki09}. Computing $\mcz(\mcc)$ is equivalent to finding the set of all consistent (i.e., path-independent) string operators for the objects in $\mcc$. Having found the set of string operators, we can then use them to see how equation (\ref{eq:onsite_twist_equation}) must be altered to account for more general phases. 
%This procedure is similar to the flux-fusion construction used to detect anomalous crystalline symmetry fractionalization employed in ref \cite{Hermele15}. In the flux-fusion test, symmetry fluxes are inserted into the theory to probe the system's ability to fractionalize the symmetry. A problem arises when the symmetry in question is that of a space group, since inserting the fluxes usually breaks the symmetry. The solution is to gauge the symmetry and study symmetry fractionalization in the gauged theory, later mapping back the knowledge of how the gauged theory behaves to determine how the symmetry fractionalizes in the pre-gauged system. %This is essentially what we're doing with the deconfinement procedure -- we run into problems when we try to study how the symmetry fluxes fractionalize, and so we make them dynamical and use their fractionalization properties in the deconfined phase to tell us about how they behaved in the confined phase. 

%One way to deconfine the symmetry fluxes in a phase $\mcc$ would be to gauge their symmetry group. The resulting phase of dynamical defects is called a twist liquid, and has been studied extensively in refs \cite{Barkeshli14,Tarantino15,Teo15}. We find it easier to use a slightly more general method for deconfining the fluxes, by studying the quasiparticle excitations of the phase $\mcc$. The quasiparticle excitations are described by the set of consistent string operators that can be written down for the theory, the set of which correspond to a mathematical object known as the Drinfeld center of $\mcc$ .

The quasiparticle excitations in $\mcz(\mcc)$ are labeled by pairs $(a,\mcr_a)$. Here, $a$ is an object (either an anyon or a symmetry defect) in $\mcc$, which encodes the fusion properties of the composite object $(a,\mcr_a)$. The second piece of data in the pair $(a,\mcr_a)$ is $\mcr_a$, which is a collection of maps called the {\it half-braiding}, which provide information about how $a$ braids with the other quasiparticle excitations in the theory. In terms of string operators, the $a$ in $(a,\mcr_a)$ gives the type of anyon created by the string operator, and $\mcr_a$ gives the different ways in which its string can satisfy path-independence. Given the fusion rules and $F$-symbols, the half-braidings can be found by finding the set of solutions to a system of nonlinear equations known as the hexagon equations (although for some of the theories considered in this paper, one must use the more general $G$-crossed consistency equations of refs.\cite{Barkeshli14,Gelaki09}). 

The principle advantages of having the half-braidings around is that they provide a way to braid worldlines around each other (see equation \ref{eq:halfbraiding}). This allows us to carry out the top arrow of the diagram (\ref{eq:onsite_twist_relation}) for objects in $\mcz(\mcc)$ without difficulty, even if the objects are non-Abelian. 
%Another benefit of working with the quasiparticle excitations in $\mcz(\mcc)$ (rather than with the objects in $\mcc$ itself) is that we don't have to keep track of the $U_a$ factors which represent the onsite fractionalization of $G$ over the worldline degrees of freedom. This is because in $\mcz(\mcc)$ the symmetry fluxes are no longer confined semi-classical objects, as they become dynamical at the quantum level, restoring the $G$ symmetry. We can also see this from the fact that the allowable string operators defined by $(a,\mcr_a)$ are always path-independent (since the strings are not observable objects), which is possible only if all the $U_a$ are trivial. 
Exploiting these facts allows us to derive an expression for $\mco_n(a)$ in terms of the twists of the quasiparticle excitations of $\mcc$, a proof of which is sketched in appendix \ref{sec:center_formula}:
\be \label{eq:onsite_twist_equation_center} \mco_n(a) = \frac{1}{\dim \mcc} \sum_{(a,\mcr_a)\in \mcz(\mcc)} \theta^n_{(a,\mcr_a)} d_{(a,\mcr_a)}, \ee
where $\dim \mcc = \sum_{a\in \mcc} d_a^2$. A similar expression has also been derived by entirely different methods in the mathematical community to compute higher Frobenius-Schur indicators\cite{Ng07,Kashina06}.

We believe that, at least for theories defined on closed manifolds, this expression for the $\mco_n(a)$ indicators applies even to topological phases with nonzero chiral central charge $c_-$, which do not admit a conventional string-net description. When we derive this expression for $\mco_n(a)$, we essentially form a fusion tree of $n$ $a$ particles and move the one on the end of the tree to the beginning. This is a physically well-defined procedure and can always be carried out in a theory without thinking about a particular string-net realization. Additionally, this expression for $\mco_n(a)$ only involves the twists of the quasiparticle excitations in the theory, which are well-defined regardless of the value of $c_-$.

%While the deconfinement procedure illustrated above is not the same thing as gauging the symmetry group $G$, we can also use the gauged theory to compute $\mco_n$ in a similar fashion. During gauging (which occurs upon decreasing the string tension term in the string-net Hamiltonian), gauge charges attach themselves to the symmetry defects, and we can similarly write down a formula for $\mco_n(\sigma)$ in terms of the twists of the flux-charge pairs $(\sigma,a)$ where $\sigma$ is one of the symmetry defects in the original theory:
%\be \label{eq:onsite_twist_equation_gauged} \mco_n(\sigma) = \frac{1}{|G|} \sum_{a\in G} \theta^n_{(\sigma,a)}.\ee

%In particular, if $\mco_n(a) \notin \mathbb{R}$ for any $n\in\ZZ$ and any $a\in\mcc$, then we can guarantee that the phase $\mcc$ breaks parity or time-reversal without needing to attempt to solve the hexagon equations. 

\section{Examples} \label{sec:examples}

In this section, we explicitly compute the invariant $\mco_n$ for a few different topological phases. We begin with Dijkgraff-Witten topological gauge theories derived from finite gauge groups \cite{Dijkgraaf90}. In these theories, the fusion rules are derived from the multiplication law in the chosen group $G$, and the $F$-symbols are given by a choice of cohomology class $\omega \in H^3(G,U(1))$. Mathematically, these theories are also equivalent to the description of symmetry defects in 2+1D bosonic symmetry protected topological phases\cite{Chen13,Levin12,Wen12,Teo15b}.
 
Because the fusion product of two objects is uniquely determined by the multiplication on $G$, all the objects must have unit quantum dimension. 
These phases break time-reversal and parity if $\omega$ is not cohomologous to $\omega^{-1}$ (for a proof, see appendix \ref{sec:braiding}), and as such we expect them to be good places to look for nontrivial $\mco_n$.

\subsection{Bosonic state with $\ZZ_N$ symmetry }
We begin by considering the simplest class of gauge theories and take $G = \ZZ_N$. The string labels in $\ZZ_N$ can either be thought of as anyons or symmetry defects, which is a particularly natural choice if $G$ corresponds to a $\ZZ_N$ rotational symmetry of the system's lattice. %(such phases could be realized on the surface of 3D topological crystalline insulators, for example). 
The fusion rule for anyons (or symmetry defects) is simply addition modulo $N$, i.e. $a \times b = [a+b]_N$, with the notation $[a]_N = a \mod N$. The representative cocycles of $H^3(\ZZ_N,{\rm U}(1))$ are well known\cite{Propitius95}, and take the form 
\be \omega(a,b,c) = e^{\frac{2\pi i p}{N^2}a(b+c - [b+c]_N)}, \ee
where $p \in \ZZ_N$ is an integer parametrizing the different cocycle generators. We note that these are essentially Chern-Simons theories at level $p$, which we see by the schematic identification $b+c - [b+c] \sim \delta A(b,c)$ for a 1-cochain gauge field $A$ and $\omega \sim \exp(\frac{2\pi i p}{N^2} A \cup \delta A)$, where $\cup$ is the cup product and $\delta$ is the coboundary operator. 

When $p=0$, $\omega$ and $\omega^{-1}$ are trivially cohomologous, implying that the theory is braided and hence cannot break parity or time-reversal, and so we restrict our attention to $p\neq0$. Since the generic $p\neq 0$ Chern-Simons term breaks parity and time-reversal, we expect these theories to be good places to look for nontrivial invariants $\mco_n$. 

The easiest way to derive $\mco_n(a)$ in this scenario is to use (\ref{eq:onsite_F_equation}) for $\mco_n(a)$ in terms of $\omega$:
\be \ba\label{eq:Zn_On} \mco_n (a) & = \delta_{[na]_N,0} \prod_{j=1}^{n-1} \omega^{-1}({a,[ja]_N,a}) \\ & = \delta_{[na]_N,0} \exp\left(-\sum_{j=1}^{n-1}\frac{2\pi i pa}{N^2}(a+ [ja]_N - [ja + a]_N)\right) \\ & = \delta_{[na]_N,0} e^{-\frac{2\pi i n p a^2}{N^2}} \ea \ee
In particular, we see that for a $\ZZ_N$ SPT phase to realize nontrivial onsite rotational symmetries at $n$-valent vertices, there must be an $n$th root of unity in $\ZZ_N$ and that if $[pa]_N = 0$, then all $\mco_n(a)$ will be trivial. 

\begin{comment}
The table below summarizes which theories possess nontrivial rotational symmetries for small $\ZZ_N$ and vertices of small valence, with $p=1$. A \checkmark means that there is some $a \in \ZZ_N$ for which $\mco_n(a) \neq 1$. 
\be
\begin{tabular}{l*{6}{c}r}
            & $n=2$ & $n=3$ & $n=4$ & $n=5$ & $n=6$ \\
\hline
$\ZZ_2$ & \checkmark & & & &\checkmark  \\
$\ZZ_3$ & & \checkmark & & & \checkmark \\
$\ZZ_4$ &\checkmark & & \checkmark & &\checkmark \\
$\ZZ_5$ & & & & \checkmark \\
$\ZZ_6$ &\checkmark & \checkmark & & & \checkmark \\
\end{tabular}
\ee
\end{comment}

We can also use this expression for the invariant $\mco_n(a)$ to derive the twists of the quasiparticles. Equation (\ref{eq:onsite_twist_relation}) implies that the twists satisfy
\be \theta_a^n =   e^{-\frac{2\pi i n p a^2}{N^2}}. \ee
This expression only tells us $\theta_a$ up to an $n$th root of unity. In particular, we are free to set 
\be \label{eq:full_twist} \theta_a = e^{-\frac{2\pi i p}{N^2}a^2}  e^{\frac{2\pi i }{N}  m a} \ee
for any integer $m\in \ZZ_N$ without contradicting (\ref{eq:onsite_twist_relation}), due to the requirement that $na$ must be a multiple of $N$ in order for $\mco_n$ to be nontrivial. 

The second exponential in equation (\ref{eq:full_twist}) is familiar as the AB phase picked up during the braiding of $m$ bosonic charges with an $a$ flux. Thus this extra phase in our expression for $\theta_a$ can be interpreted as deriving from a collection of $m$ charges attached to the flux $a$. This factor is an undetermined ``gauge'' choice precisely because the charges are part of the vacuum sector in this phase. 

When we bring the $\ZZ_N$ charges out the vacuum sector, the fluxes pair up with the charges to form flux-charge pairs, which are precisely the quasiparticle excitations of the original $\ZZ_N$ SPT phase. So we see that the quasiparticle excitations (equivalently, the string operators) are parametrized by pairs $(a,m) \in \ZZ_N \times \ZZ_N$, with twists
\be \theta_{(a,m)} = e^{-\frac{2\pi i p}{N^2}a^2}  e^{\frac{2\pi i }{N}  m a}. \ee

Since $\mco_n$ takes on complex values for these theories, they must break parity and time-reversal. To see that this is indeed the case, we consider a time-reversal transformation of the diagram giving the definition (\ref{eq:twist_def}) of the twist $\theta_a$ for some $a\in \mcc$. The time-reversal transformation $T$ flips the diagram about the horizontal access, so that the resulting diagram evaluates to $\theta^{-1}_{T(a)}$, where $T(a)$ is the image of the anyon $a$ under time-reversal. This means that if there is no $b \in \mcc$ with $\theta_b = \theta^{-1}_a$, the phase described by $\mcc$ must break time-reversal symmetry. The above examples derived from a $\ZZ_N$ gauge group with nontrivial $\mco_n$ provide easy examples of such phases, as the quasiparticle excitations of these theories all break time-reversal in this way. 

We note that if we had only focused on trivalent vertices, much of the information about the chirality of these theories would be lost. For example, the theories derived from $\ZZ_5$ have quasiparticle excitations which break time-reversal and parity, since all nonzero elements in $\ZZ_5$ have order 5, one must go to pentavalent graphs ($n=5$) before $\mco_n$ becomes complex and the symmetry breaking is detected. This illustrates a sort of emergent behavior of the anyons in the string-net graph that is not apparent in models derived from trivalent graphs.

\subsection{Bosonic state with $D_N$ symmetry for $N$ odd}
We now compute an example where the symmetry group is non-Abelian, and take $G = D_N$ to be the $N^{th}$ dihedral group. The surface states of topological crystalline insulators again provide a natural way to realize such SPT phases. 
%$D_N$ can also be written as the semidirect product $D_N \simeq \ZZ_N \rtimes \ZZ_2$, and as such represents a large class of physically interesting models. For example, we can interpret the $\ZZ_N$ component as a discrete subset of ${\rm U}(1)$ symmetry and $\ZZ_2$ as some order two internal symmetry. 
When $N$ is odd, an explicit formula for the 3-cocycle generators is known \cite{Propitius95}. To write it down, we use the isomorphism $D_N \cong \ZZ_N \rtimes \ZZ_2$ to write elements in $D_N$ as pairs $(a,\alpha)$ with $a \in \ZZ_N$, $\alpha \in \ZZ_2$, with the $\ZZ_2$ action on $\ZZ_N$ exhibited through the ``twisted'' group multiplication law
\be (a,\alpha)(b,\beta) = ([(-1)^{\beta}a + b]_N,[\alpha+\beta]_2). \ee
The 3-cocycle generators then take the form\cite{Hu13,Propitius95}
\be \ba & \omega((a,\alpha),(b,\beta),(c,\gamma)) = \exp\bigg\{ \frac{2\pi i p}{N^2} \bigg[(-1)^{\beta+\gamma}a((-1)^{\gamma}b + c \\ & \qquad - [(-1)^{\gamma}b + c]_N) + \frac{N^2}{2}\alpha\beta\gamma\bigg]\bigg\}
\ea \ee
for $p \in \ZZ_N$ which parametrizes the different possible distinct topological phases. Once again the correspondence with Chern-Simons theory is evident, with the first term in the exponential roughly identifying with $A \cup \delta A$ and the last term with $A\cup A \cup A$. 

For this example, we use (\ref{eq:onsite_F_equation_finiteG}) to compute $\mco_n$. In particular, we note that when $(a,\alpha)$ is an even element in $D_N$ (i.e., when $\alpha = 0$), we can use the results from our analysis of $\ZZ_N$ to write $\mco_n((a,0)) = \delta_{[na]_N,0} \exp(2\pi i n p a^2/N^2)$. When $\alpha = 1$, we see that $(a,\alpha)^2 = (0,0)$ and we obtain $\mco_n((a,1)) = \delta_{[n]_2,0}(-1)^{pn/2}$. We see that if $p \neq 0$ these phases must break parity and time-reversal as well. 

\subsection{Generalized Ising theories}
 
In this section we consider a simple non-Abelian theory described by the fusion algebra $\mcc = \{ \unit, \psi_1,\dots,\psi_N\} + \{\sigma\}$. They consist of a collection of Abelian anyons $\{\psi_a\}$ with $a\in\ZZ_N$, together with a single non-Abelian symmetry flux $\sigma$, which satisfy the following fusion rules:
\be\ba & \psi_a \times \psi_b = \psi_{[a+b]_N},~~\sigma \times \psi_a = \psi_a \times \sigma = \sigma,\\ &\sigma\times \sigma = \unit + \psi_1 + \dots + \psi_N.\ea\ee
This model is similar to a $\ZZ_N$ plaquette model with the defect $\sigma$ arising from lattice dislocations \cite{You12}, and although we can generically replace $\ZZ_N$ with any finite Abelian group, we will specialize to $\ZZ_N$ for simplicity. 

These theories are also naturally realized in bilayer systems, where $\sigma$ is the symmetry defect associated with the layer exchange symmetry which maps $\psi_a$ to $\psi_{[-a]_N}$. Alternatively, they can be interpreted as a set of Abelian $\ZZ_N$-anyons together with $\ZZ_2$ charge conjugation symmetry. For the quasiparticle excitations in this theory to satisfy the more general $G$-crossed consistency conditions of ref.\cite{Barkeshli14}, there must be only one fixed anyon under charge conjugation, and so we will restrict ourselves to odd $N$. 

%These theories have been of great interest in the mathematical community, where they go under the name of $\mathcal{TY}_{G}$\cite{Etingof09}. 

The most important aspect of the fusion rules involving the flux $\sigma$ is that $N^{\sigma}_{\psi_a\sigma} = N^{\sigma}_{\sigma \psi_a} = 1$, meaning that $\sigma$ localizes $N$ zero-modes, one for each particle $\psi_a$. 
%We can thus interpret this theory as having ``projective fusion rules'', since the fusion of two $\sigma$ fluxes is only well-defined up to $G$ anyons. The $G$ anyons are part of the vacuum sector, and as such can freely attach themselves to the $\sigma$ fluxes. 
This sort of behavior is characteristic of theories with non-Abelian symmetry fluxes. In this scenario, we see that the $\sigma$ defect is a generalization of a Majorana quasiparticle. 

In these theories, the $F$-symbols involving only $\psi_a$ anyons are all trivial, while the nonzero $F$-symbols involving $\sigma$ are \cite{Tambara98}
\be \ba & [F^{\psi_a\sigma \psi_b}_{\sigma}]_{\sigma\sigma} = [F^{\sigma \psi_a \sigma}_{\psi_b}]_{\sigma\sigma} = e^{\frac{2\pi i ab}{N}},\\ & [F^{\sigma\sigma\sigma}_{\sigma}]_{\psi_a\psi_b} = \frac{\alpha_{\sigma}}{\sqrt{N}} e^{-\frac{2\pi i ab}{N}},\ea\ee
where the choice of $\alpha_{\sigma} = \pm1$ parametrizes the different solutions of the theory.

%These theories are parametrized by a choice of $\alpha_{\sigma}$ and a function $\chi : G\times G \ra {\rm U}(1)$ called a symmetric bicharacter, which satisfies 
%\be \ba & \chi(a,b) = \chi(b,a),~~\chi(a,0) = \chi(0,a) = 1,\\ & \chi(ab,c) = \chi(a,c)\chi(b,c),~~a,b,c\in G. \ea \ee

We can use the first formula (\ref{eq:onsite_F_equation}) for $\mco_n$ to derive $\mco_n(\psi_a) = \delta_{a^n,1}$ for all $n$ and for all $\psi_a$. We do this by noting that the $\psi_a$ particles form an Abelian fusion algebra (allowing us to use equation \ref{eq:onsite_F_equation}) and noting that the $F$ symbols of the $\psi_a$ anyons are all trivial. From the fusion rules we see that 
$\sigma^{\times n}$ decomposes into only $\sigma$ particles if $n$ is odd, and decomposes into multiple copies of all the $\psi_a$ particles if $n$ is even. Thus we can set $\mco_n(\sigma) = 0$ if $n$ is odd (since then the vertex $|\sigma\rangle^{\tp n}$ is unstable), while $\mco_n(\sigma)$ can in general be nontrivial for any even $n$. We can then write 
\be \ba \mco_n(\sigma) = &\frac{\alpha_{\sigma}^{n/2}}{(\sqrt{N})^{n/2-1}} \sum_{u_1\in\ZZ_N} \dots \sum_{u_{\frac{n}{2}-1}\in\ZZ_N} \\ & \exp\left(\frac{2\pi i}{N} \sum_{k=1}^{n/2-1}(u_k^2 - u_ku_{k+1})\right),\ea\ee
where $u_{n/2} = 0$. Explicit evaluations of this expression are possible for fixed choices of $N$. For example, for $N = 3$, we find
\be \mco_n(\sigma) = \delta_{[n]_2,0}(\alpha_{\sigma})^{n/2} \exp\left(\frac{\pi i}{2} \left\lceil \frac{n/2-1}{3} \right\rceil\right), \ee
where $\lceil x \rceil$ denotes the smallest integer greater than or equal to $x$. As a consistency check, we note that $\mco_2(\sigma) = \alpha_{\sigma}$, as it should be. Since $\mco_n(\sigma)$ is complex, this theory must break parity or time-reversal. This is indeed the case, as $\sigma$ has no image in $\mcc$ under time-reversal. We have also checked that this result agrees with the one obtained by using \eqref{eq:onsite_twist_equation_center} and the topological twists of $\mcz(\mcc)$, which have been tabulated in the mathematical community\cite{Gelaki09}.

\subsection{3-Fermion model with $\ZZ_3$ symmetry}
We briefly show how our analysis can be applied to the example of the chiral SO$(8)_1$ state with $\ZZ_3$ symmetry. The SO$(8)_1$ model consists of three mutually semionic fermions labeled by $\psi_i$ with $i = 1,2,3$ which form a $\ZZ_2\times \ZZ_2$ fusion algebra, and is predicted to occur on the surface of certain 3D bosonic SPT states\cite{Burnell14,Teo15}. The nonzero chiral central charge of this theory leads to gapless edge modes if the theory is placed on an open manifold, and since the discussion of boundary conditions (particularly gapless ones) complicates the discussion of the $\mco_n$ indicators, we will continue to work exclusively on {\it closed} manifolds (we choose $S^2$ for concreteness). 

As noted in ref \cite{Barkeshli14}, this system possesses a natural $\ZZ_3$ symmetry action which permutes the $\psi_i$ particles, and so we are prompted to introduce $\ZZ_3$ symmetry fluxes into the theory corresponding to this permutation symmetry. There will be two such symmetry fluxes, corresponding to the two nontrivial elements of $\ZZ_3$. We denote these symmetry fluxes by $\sigma,\bar\sigma$. The fusion rules of the theory are\cite{Barkeshli14,Teo15}
\be\ba &\sigma \times \psi_i = \sigma,~~\sigma \times\bar\sigma = 0 +\psi_1+\psi_2+\psi_3,\\& \sigma\times \sigma = \bar\sigma+\bar\sigma,~~\bar\sigma\times\bar\sigma=\sigma+\sigma.\ea\ee
We note that in this case $N^{\sigma}_{\bar\sigma\bar\sigma} = N^{\bar\sigma}_{\sigma\sigma} = 2$, which means that extra vertex-based degrees of freedom need to be incorporated into the diagrammatics to account for the fusion multiplicities \cite{Lan14}, which requires a minor alternation of equation (\ref{eq:onsite_F_equation}). However, we can still apply our formula for $\mco_n$ in terms of the twists of the system's excitations to compute the onsite symmetry action. Instead of computing the full spectrum of quasiparticle excitations, we look only at the theory after gauging the $\ZZ_3$ symmetry, which suffices for our purposes since gauging deconfines $\sigma$ and $\bar \sigma$ and we can still define a consistent half-braiding for the resulting gauged theory. 

After gauging, each symmetry flux splits into three, since the original flux can absorb either $0,1,$ or $2$ units of the $\ZZ_3$ gauge charge. The twists of the resulting flux-charge pairs are \cite{Barkeshli14,Teo15}
\be \theta_{(\sigma,0)} = e^{\frac{2\pi i m}{9}},~~\theta_{(\sigma,1)} = e^{\frac{2\pi i(m+3)}{9}},~~\theta_{(\sigma,2)} = e^{\frac{2\pi i (m-3)}{9}},\ee
with the twists of the gauged $\bar\sigma$ particles identical to those of the $\sigma$ particles. Here, the integer $m \in \ZZ_3$ parametrizes the different solutions for the gauged theory. To compute $\mco_n(\sigma)$ by using the gauged theory, we need only change our normalization convention slightly, to account for the different number of flux-charge pairs in the gauged theory as opposed to the full spectrum of quasiparticle excitations. Since there are three quasiparticle species in the gauged theory associated with the flux $\sigma$, our formula for $\mco_n(\sigma)$ reads
\be \mco_n(\sigma) = \delta_{[n]_3,0}\frac{  \theta_{(\sigma,0)}^n +  \theta_{(\sigma,1)}^n +  \theta_{(\sigma,2)}^n}{3}, \ee
where the delta function is due to the fact that $\sigma$ must be fused with itself three times to yield the vacuum. So then
\be \mco_n(\sigma) = \mco_n(\bar \sigma) = \delta_{[n]_3,0} e^{\frac{2\pi i m n }{9}}. \ee
In particular, for $m = 1$ and $n = 3$ we obtain $\mco_3(\sigma) = \mco_3(\bar\sigma) = e^{2\pi i/3}$, a result which was noticed in refs. \cite{Barkeshli14,Teo15} in the form of a nontrivial $\ZZ_3$ action on the vertex $|\sigma,\sigma;\bar\sigma\rangle$. 

\section{Discussion} \label{sec:discussion}

In this paper, motivated by a recent work of Lin and Levin\cite{Lin14}, we have generalized the Levin-Wen string-net model to phases that break time-reversal and parity symmetries. In doing so, we extended the string-net model to graphs of arbitrary valence, which reveals information about certain collective properties of anyons that are obscured in conventional string-net models. A key ingredient in our generalized models was the introduction of a branch cut at each vertex into the graphical calculus, and we showed how these branch cuts naturally give rise to a certain form of onsite rotational symmetry which finds a natural motivation in our generalized tensor-product state construction. We derived formulae for the fractionalized symmetry action $\mco_n$, and showed that the phases $\mco_n$, known as higher Frobenius-Schur indicators, could be used as a computationally efficient way to detect topological phases which break time-reversal or parity. 

While our theorem 1 shows that if $\mco_n(a)\notin \mathbb{R}$ for at least one integer $n$ and $a\in\mcc$ the phase $\mcc$ must break parity and time-reversal, we are not sure whether or not the converse statement (namely, theories that break time-reversal and parity must have some complex indicator $\mco_n(a)$) is true, although we suspect that it is. One might be tempted to say that the chiral three-fermion $SO(8)_1$ model (without the $\ZZ_3$ symmetry considered in the examples section) is a counterexample, since its nonzero chiral central charge breaks time-reversal symmetry and yet $\mco_n(\psi_i) = \delta_{[n]_2,0}$ by virtue of \eqref{eq:onsite_twist_equation}, and so the $\mco_n$ are always real. However, chiral central charge is only an observable on open manifolds, and and when placed on a closed manifold (which we assumed was the case when deriving our expressions for the $\mco_n$ indicators) the $SO(8)_1$ theory {\it is} invariant under parity and time-reversal\cite{Chen15}. Understanding how to modify our calculation for the $\mco_n$ indicators for more general boundary conditions, as well as proving (or finding counterexamples to) the converse of theorem 1, would help us better understand the physics of the higher Frobenius-Schur indicators. 

Many of the examples considered in this paper can be interpreted as deriving from topological phases with symmetry defects. They are useful for us since they partially break the braiding of the ambient topological phase in which they live while still allowing for a consistent $G$-crossed theory, which is needed for the invariant $\mco_n$ to take on complex values (see appendix \ref{sec:braiding}). However, the presence of symmetry defects is not necessary for using $\mco_n$ to detect phases which break parity and time-reversal, as we have seen in several examples. As another example, it was noticed in Ref. \cite{Bonderson07} that certain parity-breaking SO$(3)_4$ models have nontrivial $\mco_3$, and we find that in general $\mco_n$ is nontrivial as long as $n$ is a multiple of 3. It would be interesting to try to find more partially braided phases like SO$(3)_4$ which are not drawn from phases with symmetry defects and to study their common properties. 

It was mentioned in the main text that if an Abelian topological order in question has a symmetry group $G$ and includes symmetry defects that induce a nontrivial $G$ action on the worldline degrees of freedom, our result for $\mco_n(a)$ in terms of the topological twist $\theta_a$ must be modified. This is because sliding worldlines under vertices as was done in the rightmost arrow in equation (\ref{eq:onsite_twist_relation}) may result in nontrivial phase factors. If $a_g$ is an Abelian symmetry defect graded by the group element $g\in G$ and we make use of the notation for the symmetry fractionalization over the worldline degrees of freedom as used in ref.\cite{Barkeshli14}, we find that
\be \mco_n(a) = \delta_{a_g^n,0}\theta_{a_g}^n \prod_{k=1}^{n-1} \eta_{a_g}(g,g^k).\ee
It is interesting to note that this expression is identical to the phase factor picked up upon winding an $a_g$ worldline $n$ times around one of the non-contractible cycles of a torus. That is, for Abelian theories we have $\mco_n(a) = [\mathcal{T}^n]_{a_ga_g}$, where $\mathcal{T}$ is the $G$-crossed $T$-matrix of the theory defined in Ref.\cite{Barkeshli14}. It would be interesting to see whether a similar interpretation holds for non-Abelian theories as well.

%While we have demonstrated that we must look to non-braided phases if we want to realize models which break parity and time-reversal symmetries, the general relationship between time-reversal and parity symmetry breaking and the ability of a system to fractionalize a rotational symmetry in the manner considered here is unclear. Indeed, it is natural to wonder whether or not only chiral phases can possess nontrivial $\mco_n$, and if this question can be answered within a more general microscopic theory of symmetry fractionalization. 

%Additionally, we can now look back at our finding that only vertices with identical outgoing anyon worldline labels can fractionalize a rotational symmetry. A rather simple (if naive) interpretation for systems with symmetry defects is as follows: normally when symmetry fluxes are inserted into the system, they explicitly break the rotational symmetry (either fully or partially), due to the fact that they are non-dynamical objects. However, if we place an identical symmetry flux at each outgoing worldline of an $n$-valent vertex, we retain a $\ZZ_n$ rotational symmetry about that vertex, and thus the rotational symmetry can still be consistently fractionalized. How exactly this argument applies to phases which possess no confined symmetry fluxes but still have nontrivial $\mco_n(a)$ for some $n$ is unclear. 

The invariants $\mco_n(a)$ were shown to be gauge-invariant under the vertex-based gauge transformation (\ref{eq:vertex_gauge_tform}), but they are not invariant under more general types of transformations relating to the quantum dimensions of the theory. In this paper, we always choose all the quantum dimensions to be real, which is physically motivated and ensures the positive definiteness of the inner product of diagrams. However, we can also achieve a mathematically consistent theory by the transformation $\mco_n(a) \mapsto \lambda(a) \mco_n(a), ~d_a \mapsto \lambda(a)d_a$ where $\lambda : \mcc\ra {\rm U}(1)$ satisfies $\lambda(ab)=\lambda(a)\lambda(b)$, which leaves the topological data invariant by virtue of equation (\ref{eq:onsite_twist_equation_center}). In Ref.\cite{Lin14}, transformations of this form were considered, which had the effect of trivializing $\mco_2(a)$ and $\mco_3(a)$ in certain scenarios (in particular, some $\ZZ_N$ theories). Although we regard imaginary $d_a$ as unphysical, it is interesting to wonder about the physical meaning of the ability to trade off phase factors between the quantum dimensions and the invariant $\mco_n$. 

One other possible direction for future work could be to extend our results to systems with $\ZZ_2$ parity symmetry, which we anticipate to be relatively straightforward. This sort of extension can be accomplished by promoting the onsite $\ZZ_N$ rotational symmetry action to a full $S_N$ symmetry action by allowing for anti-cyclic permutations of the $T$-tensor indices. This can be done relatively easily in our construction by giving the branch cut an extra $\ZZ_2$ degree of freedom (say, by coloring it red or blue) that keeps track of whether the indices of its vertex's associated $T$-tensor are indexed clockwise or counterclockwise from the branch cut. The action of $\mco_n^{(v)}$ on a clockwise vertex $v$ can then be computed quite easily as $(\mco_n^{(v)})^{-1}$. 
%Some subtleties might arise in theories whose fusion rules are derived from a non-Abelian group $G$, since the way in which general vertices are evaluated (e.g., clockwise or counterclockwise) may affect their stability. However, we note that changing the handedness of the branch cut at a vertex $v$ does not affect the Hilbert space of the system if $\mco_n^{(v)}$ is nontrivial, since changing the evaluation convention at $v$ does not affect its stability (because all of $v$'s internal and outgoing worldlines will be labeled by powers of some element $a \in G$, which always commute with each other. In any case, 
In any case, it would be valuable to obtain a more precise physical understanding for the form of symmetry fractionalization considered here and to understand possible relationships between the topological properties of string-net models and the geometry of the lattices they are defined on, with the goal of using these properties to identify a string-net condensed phase in an experimental setting. 

\acknowledgements 
We thank Zhuxi Luo, Aaron Bertram, Yang Qi, and David Aasen for helpful discussions. 

\appendix 

\section{Mapping between fusion and splitting spaces} \label{sec:more_diagrams}

In this section, we derive expressions for operators that map fusion vertices to splitting vertices and vice versa, the existence of which was key in allowing us to simplify our presentation by focusing on vertices with outgoing worldlines only.%, and whose internal structures involved only splitting vertices.

Before proceeding, we recall the graphical definition of the inner product of a splitting space with a fusion space.
Given a splitting vertex $|a,b;c\rangle$ and a fusion vertex $\langle a,b;d|$, we can form their inner product by stacking $\langle a,b;d|$ on top of $|a,b;c\rangle$:
\be \label{eq:inner_product_def} \includegraphics[scale=.85]{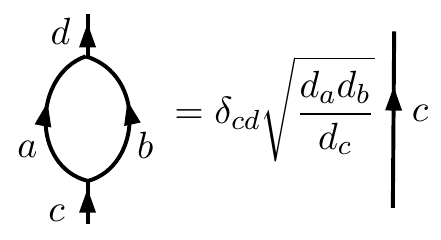} \ee
which reads $\langle a,b;d | a,b;c\rangle = \delta_{cd}\sqrt{d_ad_b/d_c} | c\rangle$. 

\begin{comment}
As mentioned earlier, the outer product gives a partition of unity through $|a\rangle \tp |b\rangle = \sum_c \sqrt{d_c/d_ad_b} | a,b;c\rangle \langle a,b;c | $, where $|a\rangle$ represents a vertical $a$ worldline. Graphically, this looks like
\be 
\includegraphics[scale=.85]{partition_of_unity.pdf}
\ee
\end{comment}

We find it helpful to introduce a second type of graph deformation similar to the $F$-move, called the $H$-move\cite{Bais14,Wen15}:
\be \includegraphics[scale=.85]{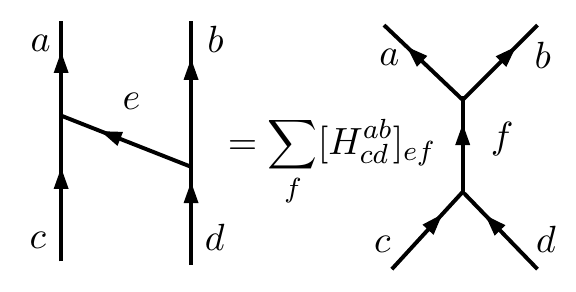}. \ee
It is related to the $F$-move through $[H^{ab}_{cd}]_{ef} = \sqrt{d_ed_f / d_ad_d} [F^{ceb}_f]^*_{ad}$, which we derive by using the unitarity of the $F$-symbols and the following diagram:
\be \includegraphics[scale=.85]{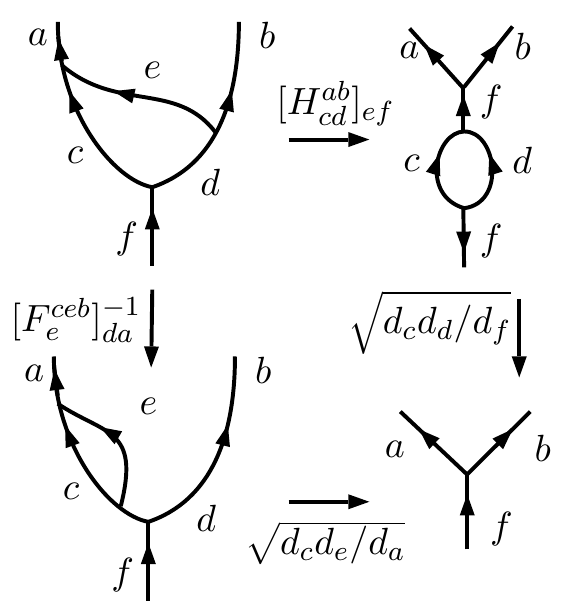}. \ee

We can use the $H$-move to derive the following operators for mapping fusion vertices to splitting vertices and vice versa:
\be \ba \label{eq:operators} & \includegraphics[scale=.85]{right_lowering_operator.pdf}, \\ & \includegraphics[scale=.85]{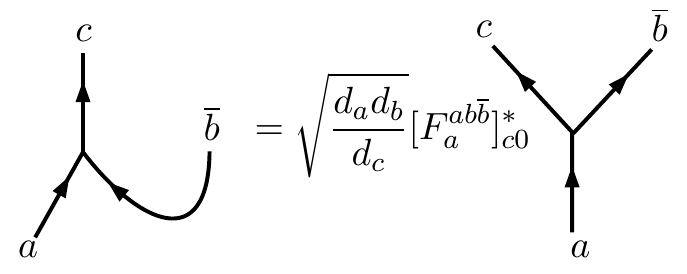}, \\ &  \includegraphics[scale=.85]{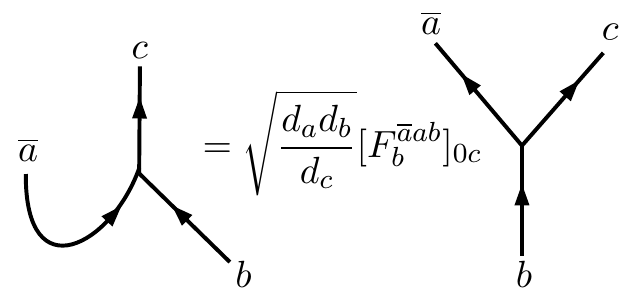}, \\ & \includegraphics[scale=.85]{left_lowering_operator.pdf}. \ea \ee
The first equality in equation (\ref{eq:operators}) can be derived by chasing down the arrow labeled by $M$ in the following diagram:
\be \includegraphics[scale=.85]{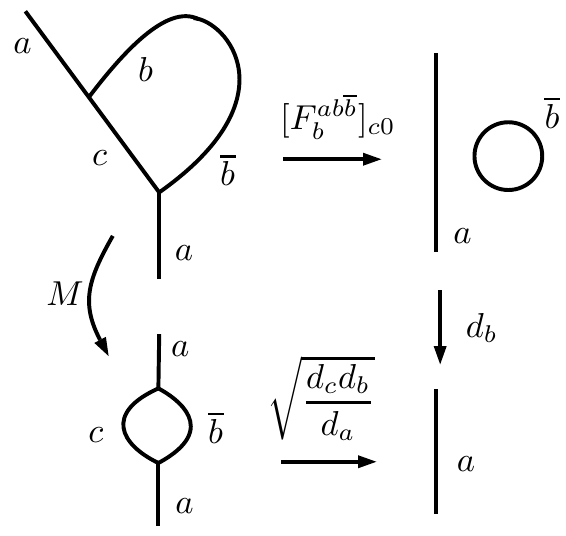} \ee
The second equality is derived by applying the $H$-move $[H^{0b}_{\overline{a}c}]_{ab}$. The third and fourth equalities are the inverses of the first two, and can be derived by using the inverted $F$-moves:
\be \ba & \includegraphics[scale=.85]{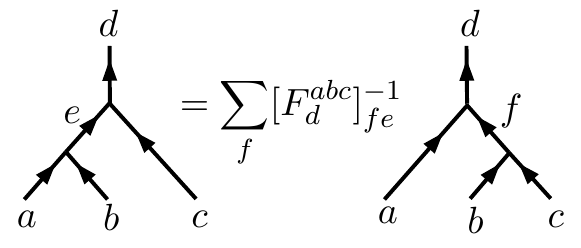}, \\ & \includegraphics[scale=.85]{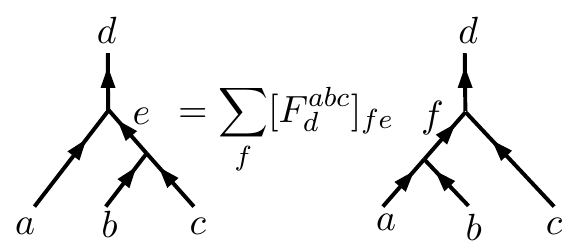}. \ea \ee

\section{Hamiltonian} \label{sec:Hamiltonian}

The Hamiltonian for our model is of the standard Levin-Wen form, modified slightly to account for our generalized string-net construction. We have
\be H = -\sum_{{\rm vertices}} A_{v_i} - \sum_{{\rm plaquettes}} B_{p_i}.\ee
The vertex projectors $A_{v}$ enforce stability at each vertex. That is, for a vertex $v = |a_1,\dots,a_n\rangle$, we have $A_v = 1$ if $0$ occurs at least once in the fusion product $a_1\times \dots \times a_n$ and $A_v = 0$ otherwise. 

The plaquette operators $B_{p_i}$ control the dynamics of the string-net graphs and are defined by
\be B_{p_i} = \frac{1}{\dim \mcc}\sum_{s\in \mcc} d_s B^s_{p_i},\ee 
with $\dim \mcc = \sum_{a\in\mcc} d^2_a$. $B^s_{p_i}$ annihilates any unstable string-net configuration, and only affects the worldline labels on the boundary of $p_i$. $B_{p_i}^s$ is computed by inserting a closed loop of $s$ string into the middle of the plaquette and then using the local rules presented in the previous appendix to fuse the loop onto the plaquette boundary\cite{Levin05}. Since fusing an $s$ loop onto a plaquette followed by an $\bar s$ loop is the same as fusing an $s\times \bar s$ loop, the plaquette operators satisfy $(B^s_{p})^{\dagger} = B^{\bar s}_p$, which implies the Herminicity of the Hamiltonian. 

To be more explicit, we demonstrate the form the plaquette operators take when acting on a square lattice. We choose a bipartite structure for the worldlines on the lattice and make the following choice for the branch cut orientations:
\be \includegraphics[scale=.75]{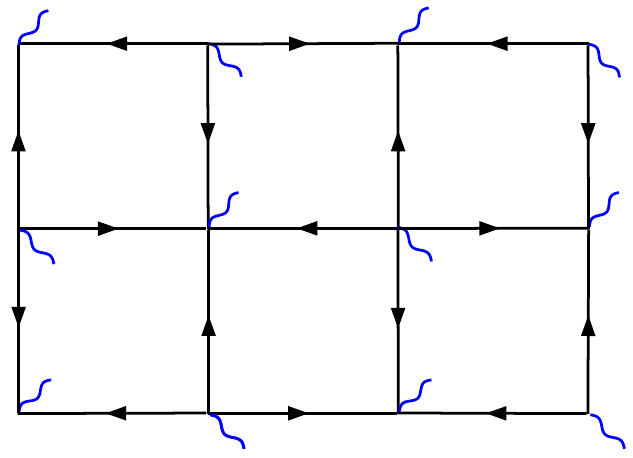} \ee

When calculating the matrix elements for $B_p^s$, we use the decomposition convention chosen in equation (\ref{eq:internal_Ttensor_def}) to decompose the tetravalent vertices into pairs of trivalent vertices, take a trace over the newly created internal degrees of freedom, and use equation (\ref{eq:onsite_F_equation}) to write out $\mco_4^{(v)}$ explicitly in terms of the $F$-symbols. Because the expressions are rather cumbersome and the general calculational framework has been discussed elsewhere\cite{Lin14,Levin05}, we will not give the details of the calculation, and merely state our result for the matrix elements of $B^s_p$:
\begin{widetext}
\be \ba  \Bigg\langle & \raisebox{-30pt}{\includegraphics[scale=.655]{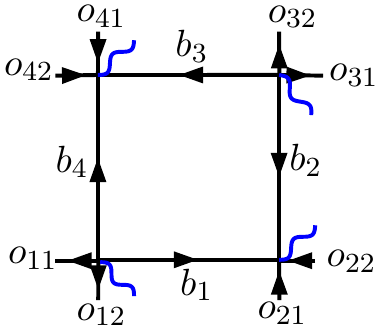}}\Bigg| B_p^s  \Bigg| \raisebox{-30pt}{\includegraphics[scale=.655]{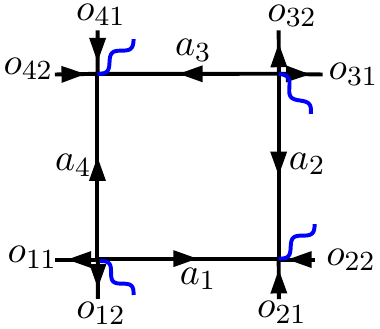}} \Bigg\rangle = \sum_{u_1,u_2,u_3,u_4 \in \mcc} N_{\bar o_{11}}^{u_1o_{12}} N^{u_2o_{21}}_{o_{22}} N^{o_{31}o_{32}}_{u_3}N^{o_{42}o_{41}}_{u_4} \frac{\alpha_{u_3}\alpha_{u_4}}{\alpha_{b_2}\alpha_{b_3}}d_{u_3}d_{u_4}d_{b_2}d_{b_3}d_s^2 \\ & \times \sqrt{\frac{d_{a_1}d_{a_2}d_{a_3}d_{a_4}}{d_{b_1}d_{b_2}d_{b_3}d_{b_4}}}  [F^{\bar u_4 u_4a_4}_{a_4}]_{0\bar a_3} [F^{u_4\bar u_4 \bar b_3}_{\bar b_3}]^*_{0b_4} [F^{a_3 u_3 \bar u_3}_{a_3}]_{\bar a_2 0} [F^{b_4 b_3 \bar b_3}_{b_4}]^*_{\bar u_4 0}[F^{a_2\bar a_2 \bar u_3}_{\bar u_3}]_{0a_3} [F^{\bar b_2 b_2 b_3}_{b_3}]^*_{0\bar u_3} [F^{\bar u_4 \bar a_3 a_3}_{\bar u_4}]_{a_4 0} [F^{\bar b_2 \bar u_3 u_3}_{\bar b_2}]^*_{b_30} \\ & \times
[F^{\bar s s a_1}_{a_1}]_{0b_1} [F^{a_2\bar s s }_{a_2}]_{b_20} [F^{\bar s s a_3}_{a_3}]_{0b_3} [F^{a_4 \bar s s}_{a_4}]_{b_40} [F^{a_4\bar s b_1}_{u_1}]_{b_1a_1}^* [F^{b_2 s a_1}_{u_2}]^*_{a_2b_1} [F^{a_2 \bar s b_3}_{\bar u_3}]^*_{b_2a_3} [F^{b_4sa_3}_{\bar u_4}]^*_{a_4b_3}. \ea\ee
\end{widetext}
In the above expression, the $\{u_i\}$ represent the internal degrees of freedom created during the decomposition of the tetravalent vertices. %In this form, the Hamiltonian is such that it commutes with the $\mco_n$ operators which rotate the branch cuts at each vertex. 

For any choice of the underlying string-net lattice structure, the vertex and plaquette operators are mutually commuting:
\be [A_{v_i},A_{v_j}] = 0,~[A_{v_i},B^s_{p_j}] = 0,~[B^s_{p_i},B^t_{p_j}]=0,\ee
which imply the exact solubility of the Hamiltonian. The first two relations are straightforward, while the third is less trivial. To check this relation, 
%we note that since fusing an $s$ string followed by a $t$ string to the same plaquette is the same as fusing an $s\times t$ string and since $s\times t = t\times s$, we see that $B_{p_i}^sB_{p_i}^t = B_{p_i}^{t}B_{p_i}^s$. Of course when the plaquettes $p_i$ and $p_j$ are not adjacent, $[B_{p_i}^s,B_{p_j}^t] = 0$, and so we just need to be concerned with the case when $p_i$ and $p_j$ represent adjacent plaquettes. 
we note that it is straightforward to check that $B_{p_i}^sB_{p_j}^t$ and $B_{p_j}^tB_{p_i}^s$ both result in sums of string-net configurations with identical edge labels. However, it is not immediately clear that the matrix elements of both operators are equal to one another. This is usually checked to be the case explicitly, although we will employ a much less tedious proof. We simply appeal to the coherence theorem in category theory\cite{Maclane78}, which in our context states that if two string-net diagrams are related to each other by two different ways of applying $F$-moves, then the two sequences of $F$-moves must have the same matrix elements (which holds since the $F$-symbols satisfy the pentagon equations). This implies that the matrix elements of $B_{p_i}^sB_{p_j}^t$ and $B_{p_j}^tB_{p_i}^s$ are indeed equal, and so the final commutation relation follows. 

\section{A proof of theorem 1} \label{sec:braiding}

Recall the definitions of the $F$-symbols and the $R$ matrix from Section \ref{sec:review} of the main text. A phase $\mcc$ is invariant under time-reversal and parity transformations if it posses a set of $F$-symbols and $R$ matrices such that the following diagram (the {\it Hexagon equations}) commutes\cite{Etingof15,Barkeshli14,Teo15,Bonderson07} (returning to the ``upward flow of time'' notation):
\be \label{eq:hex_diagram} \includegraphics[scale=.85]{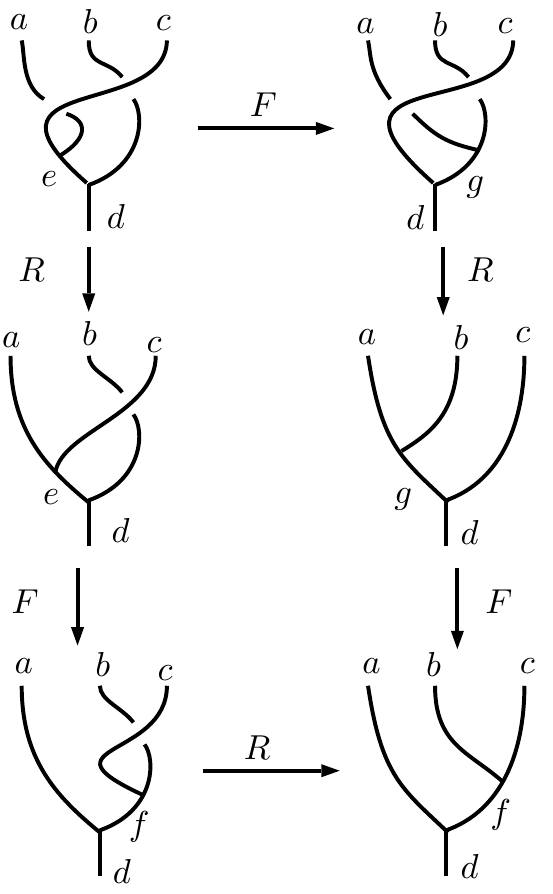}.\ee
%We emphasize that the adjective ``braided'' has nothing to do with the statistics of the quasiparticles in $\mcc$ -- it just means that it is possible to choose $\{F,R\}$ such that the above diagram commutes. 
In equation form, this is written as
\be \label{eq:hexagons} R^{ca}_e [F^{acb}_d]_{ef} R^{cb}_f = \sum_g [F^{cab}_d]_{gf} R^{cg}_d [F^{abc}_d]_{fg}.\ee
To see why theories that posses a solution to the above equations are parity and time-reversal symmetric, we observe that a set of $F$-symbols and $R$-matrices that satisfy the hexagon equations establishes a natural equivalence between $\mcc$ and the reversed topological phase $\mcc^{rev}$, where $\mcc^{rev}$ is defined as the same phase as $\mcc$, but with all diagrams reflected about the vertical axis. This is because we can use compositions of $R$-moves and $F$-moves to self-consistently relate any fusion process in $\mcc$ to the corresponding fusion process in $\mcc^{rev}$. Mathematically, this follows from the monoidal equivalence between $\mcc$ and $\mcc^{rev}$ induced by the $R$ matrices. 

If $\tau : \mcc \rightarrow \mcc^{rev}$ is the map which reflects diagrams about the vertical axis, then the equivalence between $\mcc$ and $\mcc^{rev}$ means that the image of an anyon $a\in \mcc$ can always be identified with some anyon $a_{rev} = \tau(a) \in \mcc^{rev}$. We consider what happens when we apply $\tau$ to a twisted $a$ worldline:
\be\includegraphics[scale=.85]{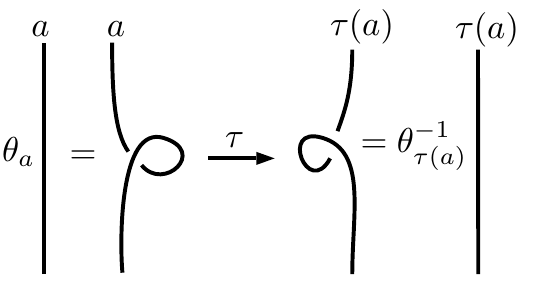}.\ee
Now $\tau(a)$ must be an anyon in $\mcc$ by the equivalence between $\mcc$ and $\mcc^{rev}$, and so for every $a\in \mcc$ there must be an anyon $\tau(a) \in \mcc$ such that $\theta_a = \theta^{-1}_{\tau(a)}$. That is, every $a$ must come with a corresponding time-reversed partner with a twist opposite to that of $a$, meaning that the phase $\mcc$ is time-reversal symmetric. An identical argument can also be applied for parity symmetry.

We will now prove our main result:
\begin{customthm}{}\label{mcotheorem} If $\mco_n(a)$ is not real for at least one choice of $n\in \ZZ$ and $a\in\mcc$, then the topological phase described by $\mcc$ must break parity and time-reversal.  \end{customthm}
This means that if $\mcc$ breaks parity and time-reversal, then it does not admit solutions to equation (\ref{eq:hexagons}). That said, $\mcc$ always allows solutions to the more general $G$-crossed consistency conditions defined in refs \cite{Barkeshli14,Etingof15}. 

We will prove the contrapositive: If $\mcc$ is parity and time-reversal invariant, then $\mco_n(a)$ must be real for all $a\in \mcc$. To avoid using excessive category theory language we will not give a mathematically rigorous proof, and will merely provide a sketch of how the proof is constructed. 
To begin, we define fusion in $\mcc^{rev}$ to be reveresed fusion in $\mcc$, so that $a \times^{rev} b = b \times a$. Now, suppose $\mcc$ is parity and time-reversal symmetric. $\mcc$ then comes with a set of $R$ matrices which allow us to construct maps $\tau :\mcc \ra \mcc^{rev}$ which flip diagrams about the vertical axis and factor over the vertices of a diagram through the $R$ matrices in a way which commutes with the $F$-moves. When performed in $\mcc^{rev}$, the onsite branch cut rotation in the computation of $\mco_n$ is also reversed, and so we have 
\be \mco_n^{rev} : T^{a_1\dots a_n} \mapsto T^{a_2\dots a_1}.\ee This means that the following diagram commutes: 
\be \label{eq:flip_square} \includegraphics[scale=.85]{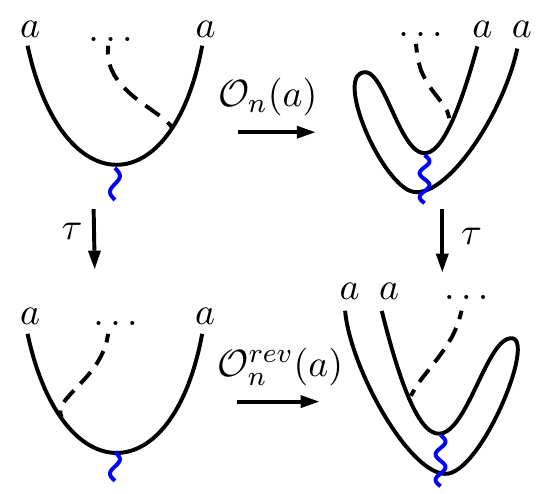}.\ee

Since the vertices of the diagrams on the top left and top right of equation (\ref{eq:flip_square}) are the same except for the two 2-valent vertices in the top-right diagram created by the worldline that is curled around the vertex, the only difference between the action of the two vertical arrows in the diagram is the action of $\tau$ over these two vertices. But this action is trivial, since the action of $\tau$ on one vertex is the inverse of its action on the other. This means that the two vertical arrows give the same phase factor. Since the diagram commutes, we have $\mco_n(a) = \mco_n^{rev}(a)$ for all $a\in \mcc$. More generally, a similar argument can be used to show that $\mco_n$ is invariant under monoidal equivalence: if $\mcf : \mcc \ra \mathcal{D}$ is a functor exhibiting a monoidal equivalence between two categories $\mcc$ and $\mathcal{D}$, then $\mco_n(\mcf(a)) = \mco_n(a)$. 

However, it turns out that the indicators also satisfy $\mco_n^{rev}(a) = (\mco_n(a))^*$ for all $n \in \ZZ$ and $a \in \mcc$. Indeed, replicating our arguments from section \ref{sec:fusiondata_mco} of the main text, we see that $\mco^{rev}_n(a)$ is given in terms of the fusion data as 
\be \mco^{rev}_n(a) = \alpha_a \sum_{\{u_i\}}\prod_{k=1}^{n-2}[F^{au_{k-1}a}_{u_{k+1}}]_{u_ku_k}.\ee
Comparing this expression to (\ref{eq:onsite_F_equation}) and using the unitarity of the $F$-symbols, we see that $\mco_n^{rev}(a) = {\mco_n(a)}^*$, as claimed. Since we also showed that $\mco_n(a) = \mco_n^{rev}(a)$, $\mco_n(a)$ must be real if $\mcc$ is invariant under time-reversal and parity.

\section{An alternate method for determining the time-reversal and parity symmetry for theories derived from finite groups}

We now provide an alternate way for determining whether or not a given topological phase derived from a finite group $G$ breaks parity or time-reversal. This question can of course be addressed by computing the invariants $\mco_n(a)$ explicitly, but here we detail a quicker approach. 

Let $\mcc$ denote the topological phase derived from the finite group $G$ with $F$-symbols given by the 3-cocycle $\omega\in H^3(G,{\rm U}(1))$. 
Suppose $\mcc$ were invariant under time-reversal and parity. Then we could define a set of $R$ matrices that satisfied the hexagon equations, and consistency between fusion and braiding would require that
\begin{comment}
and view them as functions $R^{a,b}_{ab} = R(a,b) : a \tp b \ra b \tp a$. We can also interpret $\omega$ as a map $\omega(a,b,c) : (a\tp b) \tp c \ra a \tp (b\tp c)$. The braiding on $\mcc}$ implies that the following diagram commutes \cite{Etingof15}:
\be
\begin{tikzpicture}
  \matrix (m) [matrix of math nodes,row sep=3em,column sep=10em,minimum width=2em]
  {
    c\tp (b \tp a) & (c\tp b) \tp a   \\
    c \tp (a \tp b) & (b\tp c) \tp a \\
     (a\tp b) \tp c & a \tp (b\tp c) \\};
  \path[-stealth]
    (m-1-1) edge node [left] {$\omega^{-1}(a,b,c)$} (m-2-1)
            edge node [above] {$T^n_X$} (m-1-2)
    (m-1-2) edge node [right] {$F$} (m-2-2)
    (m-2-1) edge node [left] {$\mathcal{D}(\unit,J^n)$} (m-3-1)
    (m-3-1) edge node [above] {$T^n_{F(X)}$} (m-3-2)
    (m-2-2) edge node [right] {$\mathcal{D}(\unit,J^n)$} (m-3-2)    
\end{tikzpicture}
\ee\end{comment}
\be \omega^{-1}(f,g,h)R^{g,h}_{gh}R^{f,gh}_{fgh} = \omega(f,g,h)R^{f,g}_{fg}R^{fg,h}_{fgh}. \ee
Mathematically, this is just the monoidal structure axiom for the functor $({\rm id} : \mcc \ra \mcc^{rev}, R^{ab} : a\times b \ra b \times a)$ exhibiting the monoidal equivalence between $\mcc$ and $\mcc^{rev}$, which exists due to the commutativity of the diagram (\ref{eq:hex_diagram}). This is equivalent to saying that $\omega=\omega^{-1} \delta R$, where $R$ is viewed as a function $R : G\times G \rightarrow { U}(1)$ and $\delta$ is the coboundary operator. Since $\delta R$ is a coboundary, it represents a gauge degree of freedom in $H^3(G,U(1))$, and so the cohomology classes of $\omega$ and $\omega^{-1}$ are equal. We thus see that if $\omega$ is not cohomologous to its inverse, $\mcc$ must break parity and time-reversal. 

\section{Deriving the onsite symmetry action in terms of the twists of the quasiparticle excitations} \label{sec:center_formula}

In this section, we sketch a derivation of equation (\ref{eq:onsite_twist_equation_center}) for $\mco_n$ in terms of the twists of $\mcc$'s quasiparticle excitations. We stress that this is not a mathematically rigorous proof, and is just intended to lay out the key steps. 

As mentioned in the main text, computing $\mco_n$ for a non-Abelain phase $\mcc$ is most easily done by deconfining the symmetry defects and working with $\mcc$'s quasiparticle excitations. The quasiparticle excitations are described by the set of consistent string operators that can be written down for the theory, the set of which correspond to a mathematical object known as the Drinfeld center of $\mcc$ \cite{Kassel12,Etingof05,Gelaki09}. We will write the set of quasiparticle excitations (equivalently, the set of consistent string operators) of a phase $\mcc$ as $\mcz(\mcc)$. The string operators in $\mcz(\mcc)$ are labeled by pairs $(a,\mcr_a)$. Here, $a$ is an object (either an anyon or a symmetry flux) in $\mcc$, which encodes the fusion properties of the composite object $(a,\mcr_a)$. The second piece of data in the pair $(a,\mcr_a)$ is $\mcr_a$, which is a collection of maps called the {\it half-braiding}, which provides information about how $a$ braids with the other quasiparticle excitations in the theory. Formally, $\mcr_a(b) : |a\rangle \tp |b\rangle \ra |b\rangle \tp |a\rangle$, while diagrammatically the half-braiding $\mcr_a(b)$ corresponds to exchanging the positions of an $a$ worldline and a $b$ worldline:
\be\label{eq:halfbraiding} \includegraphics[scale=.85]{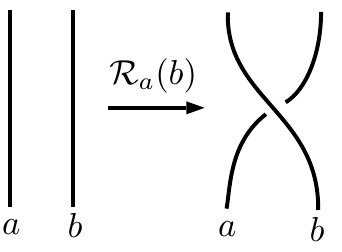}.\ee
In terms of string operators, the $a$ in $(a,\mcr_a)$ gives the type of anyon created by the string operator, and $\mcr_a$ gives the different ways in which its string can satisfy path-independence. Thus, the half-braiding maps are essentially a collection of $R$ matrices. Note that $\mcz(\mcc)$ can still admit a half-braiding structure even if $\mcc$ itself is not braided (which is true for the examples considered in this paper). 

As mentioned in the text, the half-braidings allow us to carry out the manipulation in the top arrow of the diagram (\ref{eq:onsite_twist_relation}) for objects in $\mcz(\mcc)$, even if the quasiparticle excitations in $\mcz(\mcc)$ are non-Abelian. Another benefit of working with $\mcz(\mcc)$ is that we don't have to keep track of the $\eta_{a_g}$ factors of ref.\cite{Barkeshli14} which represent the onsite fractionalization of $G$ over the worldline degrees of freedom. This is because in $\mcz(\mcc)$ the symmetry fluxes are no longer confined semi-classical objects, and so $G$ is effectively gauged in $\mcz(\mcc)$. Additionally, since strings are not physically observable, worldlines of objects in $\mcz(\mcc)$ must be path-independent, which is possible only if all the $\eta_{a_g}$ are trivial. 

To derive equation (\ref{eq:onsite_twist_equation_center}) we use the half-braiding to drag the leftmost wordline of an $n$-valent vertex under the other $n-1$ worldlines, as in the top arrow of equation (\ref{eq:onsite_twist_relation}). When we pass from $\mcc$ to $\mcz(\mcc)$, a worldline $|a\rangle$ will map to a worldline $|(a,\mcr^{(i)}_a)\rangle$, where $\mcr^{(i)} \in \{\mcr_a^{(1)},\dots,\mcr_a^{(m)}\}$ is a half-braiding for $a$ drawn from the set of all allowable half-braidings. Suppose the vertex $|a,\dots,a\rangle$ is mapped to the vertex $|(a,\mcr^{(1)}_a),\dots,(a,\mcr^{(n)}_a)\rangle$ during the transition from $\mcc$ to $\mcz(\mcc)$. Then we can compute the first step in equation (\ref{eq:onsite_twist_relation}) as follows:
\be \includegraphics[scale=.85]{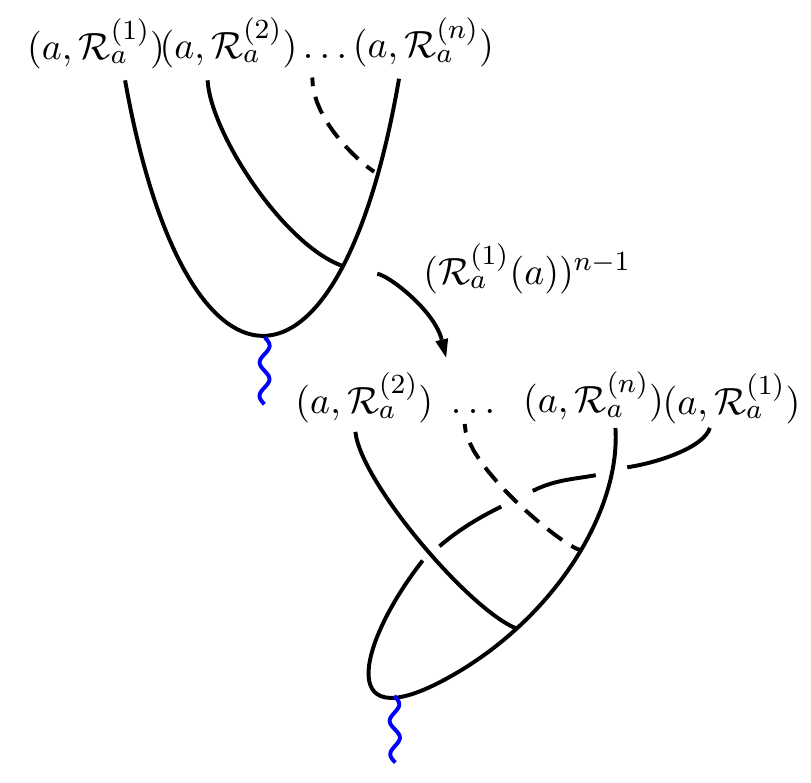}.\ee
Note that this process only depends on the choice of $\mcr_a^{(1)}$. By applying the half-braiding operation to a diagram with a twisted loop (equation \ref{eq:twist_def}) we see that $\mcr_a(a) = \theta_{(a,\mcr_a)}$ (up to a factor of $d_{(a,\mcr^{(1)}_a)}$), and so $(\mcr_a^{(1)}(a))^{n-1} \propto \theta_{(a,\mcr^{(1)}_a)}^{n-1}$. The bottom horizontal arrow of (\ref{eq:onsite_twist_relation}) gives $\theta_{(a,\mcr^{(1)}_a)}$ as before. 

This line of reasoning gives us $\mco((a,\mcr_a))$, but we do not yet know $\mco_n(a)$. As mentioned earlier, there is an ambiguity that arises when we pass from $\mcc$ to $\mcz(\mcc)$, since we can't control which $\mcr_a^{(i)}$ gets chosen at each leg of the vertex (i.e., we cannot control how many gauge charges attach themselves to $a$ during the $\mcc \ra \mcz(\mcc)$ transition).
In order to remedy this ambiguity, we simply average over all the possible choices for the half-braiding. That is, we average over all possible images of $a$ under the map $\mcc \ra \mcz(\mcc)$. 
%The correct normalization factor for the average turns out to be $1 / \dim \mcc$, where $\dim \mcc = {\sum_{a\in \mcc} d_a^2}$. 
Putting everything together, we arrive at 
\be \label{eq:onsite_twist_nonabelian_equation} \mco_n(a) = \frac{1}{\dim \mcc} \sum_{(a,\mcr_a)\in \mcz(\mcc)} \theta^n_{(a,\mcr_a)} d_{(a,\mcr_a)}. \ee
We note that this expression is essentially the same as the formula for the $n^{th}$ higher Frobenius-Schur indicator of $a$, which has been derived in the mathematical community by very different methods \cite{Ng07,Shimizu11}. 

\section{Tetrahedral symmetry breaking} \label{sec:tetra}
By putting together the diagrams on both sides of equation (\ref{eq:F_move}), we see that the $F$-move has the structure of a tetrahedron:
\be
\includegraphics[scale=.85]{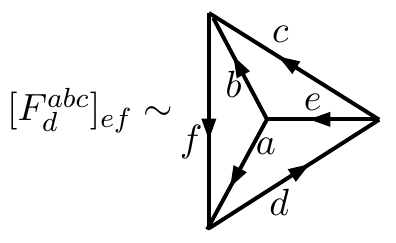}
\ee
It is often assumed that the $F$-symbols are symmetric with respect to rotations and reflections of the tetrahedron. This implies the {\it tetrahedral symmetry} of the $F$-symbols\cite{Levin05,Lan14,Hu15}:
\be [F^{abc}_{d}]_{ef} = [F^{b\overline{e} d}_{f}]_{\overline{a} c} = [F^{c\overline{d}a}_{\overline{b}}]_{\overline{e}\overline{n}} = [F^{e \overline{b} f}_{d}]_{ac}. \ee
The equalities correspond respectively to the order 3 rotation, order 2 rotation, and reflection of the tetrahedron, which collectively generate the tetrahedral group. 

In our ``upward flow of time'' convention where the arrows are left implicit in the diagrams, there is no reason to expect the regular tetrahedral symmetry relations to hold. For example, a generic tetrahedral transformation of the $F$-symbols can lead to the reversal of the arrows on some of the $F$-symbol's worldlines, which will in general be nontrivial since the models we have considered with complex $\mco_n$ also break TRS. 

We can use the local rules presented in section \ref{sec:review} to explicitly compute the action of the tetrahedral symmetry in terms of the fusion data. This is of great practical use, since knowledge of the tetrahedral symmetry relations greatly simplify the task of numerically determining the fusion data by solving the pentagon equations for generalized string-net models which are not fully braided.  

To facilitate the calculation of the generalized tetrahedral symmetry relations, we notice that $F$-symbols of the form $[F^{abc}_{0}]_{ef}$ are proportional to $\delta_{e,\overline{c}}\delta_{f,\overline{a}}$. In particular, this means that $[F^{abc}_{0}]_{\overline{c}\overline{a}}$ is just a number. To simplify the notation a bit, we will define $\zeta^{abc} = [F^{abc}_{0}]_{\overline{c}\overline{a}}$. 

The tetrahedral symmetry action can be derived by straightforward diagrammatic manipulations, using the local rules presented in section \ref{sec:review}. The general idea is to use the local rules perform mutate the tetrahedron corresponding to the original $F$-symbol into the tetrahedron corresponding to the transformed $F$-symbol. For computations, we find it easier to write the tetrahedron by putting it on a sphere and using the ``all worldlines oriented upwards'' convention, in which it looks like
\be \includegraphics[scale=.85]{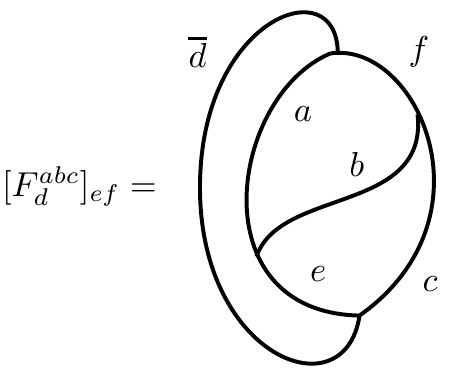}.\ee 
We will only show the derivation of one of the identities of tetrahedral symmetry (the order 3 rotation) for the sake of presentation. The computation proceeds as follows:
\begin{widetext}
\be \includegraphics[scale=.85]{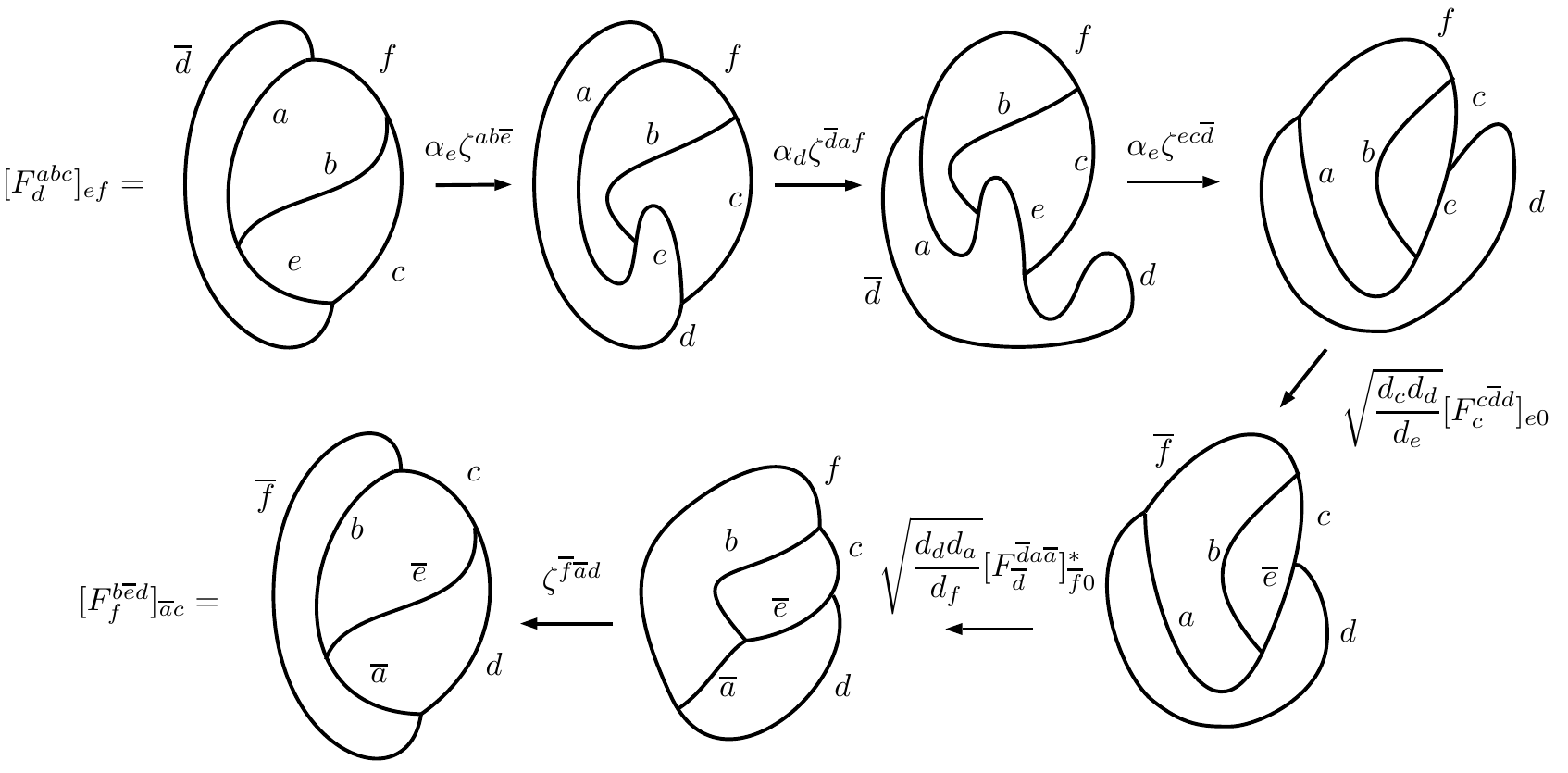} \ee
The other relations are derived by following a similar diagrammatic approach. We thus obtain the generalized relations of tetrahedral symmetry:
\begin{eqnarray} && [F^{abc}_{d}]_{ef}  = %\frac{\alpha_d \zeta^{\overline{d}af}\zeta^{ab\overline{e}} \zeta^{\overline{f}\overline{a}d} [F^{c\overline{d}d}_c]_{\overline{e}0}}{[F^{\overline{d}a\overline{a}}_{\overline{d}}]_{0c}}[F^{b\overline{e} d}_{f}]_{\overline{a} c}, \\ 
\left(\alpha_d d_d \sqrt{\frac{d_ad_c}{d_ed_f}} \zeta^{\bar d a f}\zeta^{ab\bar e}\zeta^{ec\bar d} \zeta^{\bar f \bar a d} [F^{\bar d a \bar a}_{\bar d}]^*_{\bar f 0} [ F^{c\bar d d}_{c}]_{e0}\right) [F^{b\overline{e} d}_{f}]_{\overline{a} c}, \\
&& [F^{abc}_{d}]_{ef}  = \frac{\zeta^{b\overline{e}a}\zeta^{ab\overline{e}}\zeta^{ec\overline{d}}}{\zeta^{af\overline{d}}\zeta^{e\overline{d}a}\zeta^{bc\overline{e}}} [F^{c\overline{d}a}_{\overline{b}}]_{\overline{e}\overline{f}}, \\
&& [F^{abc}_{d}]_{ef}  =  \left(\alpha_b[F^{\overline{b}bc}_c]_{0f}[F^{ab\overline{b}}_a]_{e0}\right)[F^{e \overline{b} f}_{d}]_{ac},
\end{eqnarray}
which correspond to the order 3 rotation, order 2 rotation, and reflection, respectively.
\end{widetext}

\bibliography{onsite_bibliography}

\end{document}